\documentclass[%
 reprint,
 amsmath,amssymb,
 aps,
]{revtex4-2}

\usepackage{graphicx}
\usepackage{dcolumn}
\usepackage{bm}
\usepackage[dvipsnames]{xcolor}
\usepackage{ulem}
\usepackage{braket}
\usepackage{booktabs}
\usepackage{pifont}
\usepackage{amssymb}
\usepackage{adjustbox}

\begin{document}
\newcommand{\vP}{\mathbf{P}}
\newcommand{\vers}{\mathbf{r''}}
\newcommand{\vRs}{\mathbf{R''}}
\newcommand{\verp}{\mathbf{r'}}
\newcommand{\vRp}{\mathbf{R'}}
\newcommand{\Wcm}{\;\mathrm{W/cm}^2}
\newcommand{\eV}{\;\mathrm{eV}}
\newcommand{\Rep}{\mathrm{Re}\,}
\newcommand{\Imp}{\mathrm{Im}\,}
\newcommand{\vk}{{\mathbf{k}}}
\newcommand{\vkst}{\mathbf{k}_\textrm{st}}
\newcommand{\vqst}{\mathbf{q}_\textrm{st}}
\newcommand{\vi}{\hat{\mathbf{i}}}
\newcommand{\vj}{\hat{\mathbf{j}}}
\newcommand{\vS}{{\mathbf S}}
\newcommand{\vH}{\mathbf{H}}
\newcommand{\vv}{\mathbf{v}}
\newcommand{\ve}{\hat{\mathbf{e}}}
\newcommand{\0}{\mathbf{0}}
\newcommand{\vE}{\mathbf{E}}
\newcommand{\vA}{\mathbf{A}}
\newcommand{\vT}{\mathbf{T}}
\newcommand{\ver}{\mathbf{r}}
\newcommand{\vd}{\mathbf{d}}
\newcommand{\va}{\mathbf{a}}
\newcommand{\vD}{\mathbf{D}}
\newcommand{\vp}{\mathbf{p}}
\newcommand{\vR}{\mathbf{R}}
\newcommand{\vq}{\mathbf{q}}
\newcommand{\vrho}{\mbox{\boldmath{$\rho$}}}
\newcommand{\del}{\mbox{\boldmath{$\nabla$}}}
\newcommand{\valpha}{\mbox{\boldmath{$\alpha$}}}
\newcommand{\vRR}{\{\vR\}}
\newcommand{\vRRp}{\{\vRp\}}
\newcommand{\Ip}{I_\mathrm{p}}
\newcommand{\Up}{U_\mathrm{p}}
\newcommand{\calT}{\mathbf{\cal T}}
\newcommand{\calF}{\mathbf{\cal F}}
\newcommand{\calU}{\mathbf{\cal U}}
\newcommand{\et}{\tilde{e}}
\newcommand{\cm}{\mathrm{c.m.}}
\newcommand{\vro}{\mathbf{\rho}}
\newcommand{\tos}{t_{0s}}
\newcommand{\ts}{t_{s}}
\newcommand{\Ep}{E_{\mathbf{p}}}
\newcommand{\field}[1]{\noindent\textbf{#1}\ }
\preprint{APS/123-QED}

\title{Physics-informed neural networks for solving saddle-point equations in strong-field physics with tailored fields}

\author{Jiakang Chen}
\thanks{These authors contributed equally to this work.}

\author{Sufia Hashim}
\thanks{These authors contributed equally to this work.}

\author{Carla Figueira de Morisson Faria}
\affiliation{University College London, Gower Street, WC1E 6BT, London, UK}

\date{\today}

\begin{abstract}

We develop an unsupervised physics-informed neural network (PINN) to solve the saddle-point equations (SPEs) governing direct above-threshold ionization (ATI) within the strong-field approximation. This setting provides a well-understood testbed in which the saddle-point structure is known for tailored driving fields, enabling systematic validation of the proposed solver. In this approach, the network is trained by minimizing the residual of the saddle-point equations and requires only the definition of the driving-field shape and its parameters, such as intensity, carrier-envelope phase, ellipticity, and relative phase. We introduce a window parametrization strategy that maps unconstrained network outputs to prescribed regions of the complex-time plane, guiding the optimization toward physically relevant solutions and improving convergence stability. We benchmark the PINN against a traditional local gradient-based solver for a range of monochromatic, few-cycle, bichromatic, elliptical and bicircular fields, demonstrating robust recovery of the dominant complex ionization times over wide parameter ranges. The network naturally tracks changes in ionization event dominance as laser parameters are varied, enabling continuous exploration of regimes where conventional solvers require repeated manual initialization. Using the PINN-derived solutions, we compute coherent ATI photoelectron momentum distributions and show that the symmetries of the driving fields are reflected in both the saddle-point structure and the resulting spectra. These results establish PINNs as a promising computational framework for semiclassical strong-field calculations and provide a foundation for extending machine-learning solvers to Coulomb-corrected models or to more complex strong-field processes, such as rescattered ATI for which the SPEs are highly nonlinear and the presence of multiple closely-spaced solutions makes conventional Newton-type root-finding highly sensitive to initial guesses, which hinders systematic investigations across laser-parameter spaces, particularly for tailored fields. 
\end{abstract}

\maketitle

\section{Introduction}\label{sec:intro}

A central challenge throughout theoretical physics is the evaluation of highly oscillatory integrals of the form $\int e^{iS(q)}dq$\footnote{We use atomic units throughout this work, such that $\hbar=1$.} which arise from path-integral formulations of quantum mechanics \cite{Rosenfelder2017} and quantum field theory \cite{Sanjeev2000}. When the action $S$ is large, the dominant contributions to such integrals comes from the trajectories at which the action is stationary i.e. the saddle points. 
In quantum mechanics, the Wenzell-Kramers-Brillouin (WKB) approximation and its higher-order extensions reduce tunneling problems to the evaluation of classical paths in complex time \cite{Misumi2025}. In quantum field theory, instantons (defined as finite-action saddle points of the path integral) determine, for example, vacuum tunneling in quantum chromodynamics \cite{Anders2016} and baryon number violation in electroweak sector \cite{Elder2025}. In statistical mechanics, saddle points of the Hubbard–Stratonovich transformed partition function play a significant role in generating mean field theories \cite{Altland2010}. 
In each case, the physics is incorporated in the solution of a set of nonlinear, often complex-valued, equations whose structure reflects the symmetries and topology of the underlying theory.

Saddle-point methods are also integral to strong-field physics, 
where matter is exposed to an intense laser field, comparable to the target's binding forces. This is a consequence of two main observations. First, the extreme conditions involved preclude the application of perturbation theory with regard to the field. Second, strong-field phenomena such as above-threshold ionization (ATI) and high-order harmonic generation (HHG), can be explained as resulting from the laser-induced recombination or recollision of an electron with its parent ion \cite{Corkum1993}. This physical picture is known as the `three-step model' and has been first introduced classically, in what is referred to as `the simple-man model'. Quantum mechanical formulations thereof were developed by applying saddle-point methods to quantum mechanical transition amplitudes, which are expressed as nested, rapidly oscillating integrals \cite{Salieres2001}. This has led to a myriad orbit-based methods such as the strong-field approximation (SFA), Eikonal Volkov approximation \cite{Smirnova2007JPhysB}, Analytic R-Matrix Theory \cite{Torlina2012a, Morishita2008}, Quantitative Rescattering Theory \cite{Lin2018}, Coulomb-Volkov Approximation \cite{Arbo2008, Cavaliere1980}, Quantum Monte Carlo method \cite{Geng2014}, Coulomb Quantum-Orbit Strong Field Approximation \cite{Lai2015, Maxwell2017}, Semiclassical Two-Step Model \cite{Lein2016}.
These methods associate quantum pathways to electron trajectories, offering an intuitive framework for identifying and distinguishing the underlying physical mechanisms, and for interpreting quantum interference effects that are often obscured in the full wavefunction description. When combined with tailored driving fields (such as two-color or polarization-shaped pulses \cite{Fu2012,Huang2018,Quan2009,Zhang2014,Mancuso2016,Song2018,Pang2020,Ge2023,Liu2024} this framework allows selective control over ionization pathways \cite{Chen2017}, recollision dynamics \cite{Fu2012,Zhang2014,Mancuso2016}, and electron correlation effects \cite{Hashim2024,Hashim2024b,Hashim2024c,Hashim2025}, providing a powerful interpretational tool. For an overview see \cite{Faria2020, Symphony}.  Undoubtedly the most widespread of these approaches is the SFA, or Keldysh-Faisal-Reiss (KFR) theory \cite{Keldysh1965,Faisal1973,Reiss1980}, which provides a standard semiclassical treatment and is a computationally efficient alternative to the full time-dependent Schr\"odinger equation. The SFA neglects the Coulomb interaction during the electron's continuum propagation and the laser field when the electron is bound.
These approximations often exclude important physics. However, they render the framework easily extendable to otherwise demanding scenarios, such as quantum light, or correlated multielectron dynamics, as the continuum can be described analytically with Volkov waves. For overviews see \cite{Faria2020, Symphony}.  

Solving SPEs using numerical methods remains an active research field \cite{Liu2022,Nataf2024,Kazantcev2020}, including the highly nonlinear SPEs within strong-field physics \cite{Weber2025}. The most traditional approach to solve the SPEs consists of using local gradient-based root-finding algorithms \cite{Nocedal2006}, commonly Newton-type methods \cite{Murray2011}. The equations are defined in the complex time plane and generally allow multiple closely-spaced solutions whose number, location, and relative importance depend sensitively on the driving-field parameters. Hence, this poses several challenges.

First, local optimization algorithms depend heavily on heuristic initial guesses. While effective for simple field configurations, these approaches fail when solutions move, merge, or bifurcate as laser parameters such as intensity, carrier-envelope phase, pulse duration, or polarization are varied \cite{Faria2002,Habibovic2021,Weber2025}. This lack of robustness is particularly severe near classical cutoffs and Stokes transitions \cite{Berry1989,Koch2018,Koch2018b}, where the Jacobian becomes ill-conditioned and convergence is not guaranteed \cite{Murray2011, Nocedal2006}. As a result, solvers may converge to unintended or unphysical roots, exhibit discontinuous ``jumping” between roots, or fail to converge altogether. These issues are exacerbated for tailored fields and large parameter scans, where heuristic initialization strategies become impractical and computationally expensive as iterative methods necessitate solving the SPEs at each grid point. A further challenge lies in the classification of the resulting solutions, as mathematically valid saddle points may correspond to causality-violating \cite{Shaaran2012}, divergent or exponentially suppressed transition amplitudes that either give spurious results, or contribute negligibly to observables \cite{Faria2002}. For orbit-based methods beyond the SFA, the presence of the binding potential will introduce root cuts \cite{Popruzhenko2014b,Maxwell2018b}, additional orbits \cite{Yan2010,Yan2012,Maxwell2017,Rodriguez2023}, singularities, caustics and potential-dependent symmetry breaking (see, e.g., \cite{Yan2010,Xia2018,Lopez2019b,Liao2022,Rook2024}). 

Moreover, these approaches become computationally prohibitive for large parameter sweeps, which are required for modeling focal averaging \cite{Maxwell2016},  or integrations over multiple field parameters. For example, recent studies investigating the effect of quantum light driven strong-field processes \cite{RiveraDean2025,Stammer2024} often involve solving the SPEs for many different intensities for a fixed quantum state of light and a classical field, followed by a weighted sum according to the photon number distribution of the quantum field \cite{Lyu2025, Liu2025, Habibovic2025ql}. While symmetry considerations can substantially reduce this cost for highly symmetric fields by relating solutions across optical cycles, such simplifications generally break down for tailored fields, including bichromatic or few-cycle pulses, where symmetry-related trajectories may not exist or may contribute unequally. In these cases, brute-force root-finding yields large numbers of saddle points \cite{Habibovic2025,Habibovic2025iii} whose classification and physical relevance must be determined \textit{a posteriori}, rendering local, gradient-based solvers inefficient and difficult to generalize. These limitations motivate the development of more general, robust strategies for identifying and tracking physically relevant saddle-point solutions across complex field configurations.

In the last decade, Machine learning (ML) has demonstrated the ability to accelerate or even replace conventional numerical simulators and solvers, most notably within the fields of computational fluid dynamics \cite{Caron2025}, electromagnetism \cite{Sagar2021} and quantum physics \cite{Cho2024}. Within strong-field and attosecond physics, most existing applications focus on post-processing or classification using convolutional neural networks and dimensionality-reduction methods \cite{Hajivassiliou2023,Hirschman2025,Chomet2022,Liu2020ml}. Deep-learning applications range from laser-pulse characterization \cite{Meng2023}, molecular bond-length retrieval \cite{Lein2024}, and the prediction of two-color pulses \cite{Alaa2024}, to strong-field dynamics, where neural networks are trained to reproduce observables derived from Feynman path-integral calculations \cite{Liu2020ml}. In this approach, the network learns mappings of the system parameters and final spectra, enabling prediction of strong-field observables once the model is trained. A comparatively underexplored direction is the use of ML to solve the theoretical equations governing strong-field processes themselves.

Physics-informed neural networks (PINN), \cite{Raissi2019} provide a particularly natural framework by which to tackle this problem, as they can incorporate physics information (such as the SPEs, laser field parameters and symmetry considerations) when training. Unsupervised PINNs do not require precomputed training data and can reduce the sensitivity to initial conditions that limits traditional root-finding algorithms. Moreover, their ability to generalize across parameter space enables efficient exploration of wide ranges of laser parameters, such as intensity, carrier-envelope phase, and pulse shape, distinguishing them from interpolation-based approaches and making them well suited for systematic studies of tailored strong-field dynamics.

Here, we introduce a physics-informed neural network (PINN) solver for the saddle-point equations of direct above-threshold ionization (ATI) within the strong-field approximation. It is meant as a proof of concept in which we apply these methods to the simplest possible scenario, which results in a one-dimensional integral solved by saddle-point methods. Thus, even for tailored fields \cite{Eberly1991, Ehlotzky2001, Milos2006, Bashkansky1987, Becker2017plane}, the saddle-point solutions are comparatively simple to obtain and well understood, enabling systematic validation of the method and direct comparison with established approaches. Furthermore, for direct ATI the saddles are typically well separated, while closely spaced or coalescing saddle points are more common for strong-field phenomena with recombination or rescattering. 

In addition, we approach the problem from a symmetry perspective. In strong-field laser-matter interaction, dynamical symmetries have proven particularly useful \cite{Busuladvic2017,Yue2020, Neufeld2019} over the past three decades to derive selection rules for HHG  \cite{Alon1998,Milos2015,Liu2016,Neufeld2019,Yue2020} and ATI \cite{Milosevic2016,Busuladvic2017,Habibovic2020}, to determine the shape of photoelectron momentum distributions \cite{Habibovic2021}, and, recently, to explain how different scattering properties of a soft-core and a Coulomb potential manifest themselves in the rescattered ATI spectra \cite{Rook2024}. However, the development of a more complete theory, focused on group-theoretical methods and exploring structured light of increasing complexity, as well as other degrees of freedom such as spin, angular momenta is still work in progress (see the perspective articles \cite{Habibovic2024,Neufeld2025}).

We demonstrate that unsupervised PINNs can accurately solve the saddle-point equations and generate physically meaningful complex ionization times across a wide range of laser intensities and field parameters, including carrier-envelope phase, relative phase and ellipticity. The central aim of this work is to assess whether physics-informed architectures can reproduce existing results obtained with local gradient-based methods for fields of different symmetries.

To this end, we apply the PINN solver to linearly polarized monochromatic, bichromatic, and polychromatic fields, as well as to elliptically polarized and bicircular driving fields. Although the residual Coulomb potential is not considered here, the approach is readily extendable through the inclusion of additional physical constraints. These results establish a foundation for physics-informed, data-driven solvers in strong-field physics and represent a first step toward addressing the computational bottleneck associated with identifying physically relevant saddle-point solutions in more complex ionization processes.

This article is organized as follows. In Sec.~\ref{sec:framework}, we first review the strong-field approximation transition amplitude and saddle-point equations for direct above-threshold ionization. Section ~\ref{sec:fields} introduces the specific tailored fields employed in this study along with their symmetry considerations. We recount some issues of existing solving methods before introducing the machine learning solver in Sec.~\ref{sec:impl} wherein we also discuss the training protocol, the specific implementation details such as the window parametrization method used to select the valid saddle point solutions, and provide model optimization studies. Subsequently, the direct ATI saddle point solutions and photoelectron momentum distributions are presented in Sec.~\ref{sec:results} for each of the fields. We discuss how the symmetry of the fields maps to the saddle-point solution topology and ATI distributions. Finally, in Sec.~\ref{sec:conclusions}, we state our conclusions. 

\section{Physical framework: direct ATI}\label{sec:framework}

In this section we briefly review the strong-field approximation for direct above-threshold ionization (D-ATI). 
D-ATI is one of the simplest strong-field ionization processes in which an electron tunnel ionizes from a bound state into the continuum and is subsequently driven solely by the laser field without revisiting the parent ion. Within the SFA, the Coulomb attraction after ionization can be neglected. In this framework, the ionization process can be interpreted in terms of the trajectory of the electron, where the electron is assumed to tunnel into the continuum at some well-defined time within the laser cycle, and subsequently evolve under the influence of the external laser field before reaching the detector. These trajectories should not be understood as classical paths in real space, but rather as quantum-mechanical contributions to the transition amplitude, each associated with a distinct ionization time and accumulated phase. 

Therefore the transition amplitude for D-ATI reads, 
\begin{equation}
    M(\vp) = -i \int_{-\infty}^{\infty} \! dt \, \bra{\vp + \vA(t)} \vE(t)\cdot\ver \ket{\psi_0}\,
    \exp\!\left[i S(\vp,t)\right],
    \label{eq:datiM}
\end{equation}
where $S(\textbf{p},t)$ is the semi-classical action given by
\begin{equation}
S(\textbf{p},t) = I_p t - \frac{1}{2}\int^\infty_t \text{d}\tau [\textbf{p}+\textbf{A}(\tau)]^2,
\label{eq:datiS}
\end{equation}
$I_p$ is the ionization potential and $\textbf{p}$ is the final momentum of an electron which ionized at time $t$, $\textbf{A}(t)$ is the vector potential of the driving laser field and
\begin{equation}
V_{\textbf{p}0} = \bra{\vp + \vA(t)} \vE(t)\cdot\ver \ket{\psi_0}
    \label{eq:datiprefactor}
\end{equation}
refers to prefactor associated with ionization of the electron from a bound state $\ket{\psi_0}$ into a field-dressed continuum (Volkov) state $\ket{\textbf{p} + \textbf{A}(t)}$. Note that this is written in the length gauge where ionization is described by the dipole interaction $\textbf{E}(t)\cdot\textbf{r}$. The dominant contributions to the transition amplitude arise from stationary points of the action, corresponding to ionization times for which the rapidly oscillating phase (action) becomes stationary. These times are determined by the saddle-point equation (SPE),
\begin{equation}
    \frac{\partial S(\vp,t)}{\partial t} = 
    \frac{\left[\vp + \vA(t)\right]^2}{2} + \Ip = 0,
    \label{eq:ati_spe}
\end{equation}
whose complex-valued solutions correspond to the ionization time, and define the quantum trajectories contributing to the D-ATI process.  This SPE is associated with conservation of energy of the electron at the ionization time. It is nonlinear and can allow multiple solutions per field cycle. In the limit $\Ip \rightarrow 0$, the classical ionization times obtained for the simple-man model are recovered. Although Eq.~\eqref{eq:ati_spe} has no classical counterpart, there will be a maximal kinetic energy that, classically, can be acquired from the field. This is known as the direct ATI cutoff energy \cite{Becker2002Review}, and it is proportional to the driving-field intensity. This implies that ATI photoelectron momentum distributions (PMDs) should become broader and extend towards higher momenta for increasing driving-field intensities.

In this study, we use Helium as a target, taking the initial bound-state to be the $1s$ state with an ionization potential $I_p= 0.904 \text{a.u.}$ We neglect the influence of the prefactor, as it primarily imprints the signature of the target onto the ATI distributions and introduces a momentum bias that may obscure the field-dependent effects we wish to explore in this study. This allows us to isolate the role of the tailored field and its symmetry in shaping the PMDs. 

\section{Tailored fields and symmetries}\label{sec:fields}

\begin{figure*}[t]
    \centering
    \includegraphics[width=\textwidth]{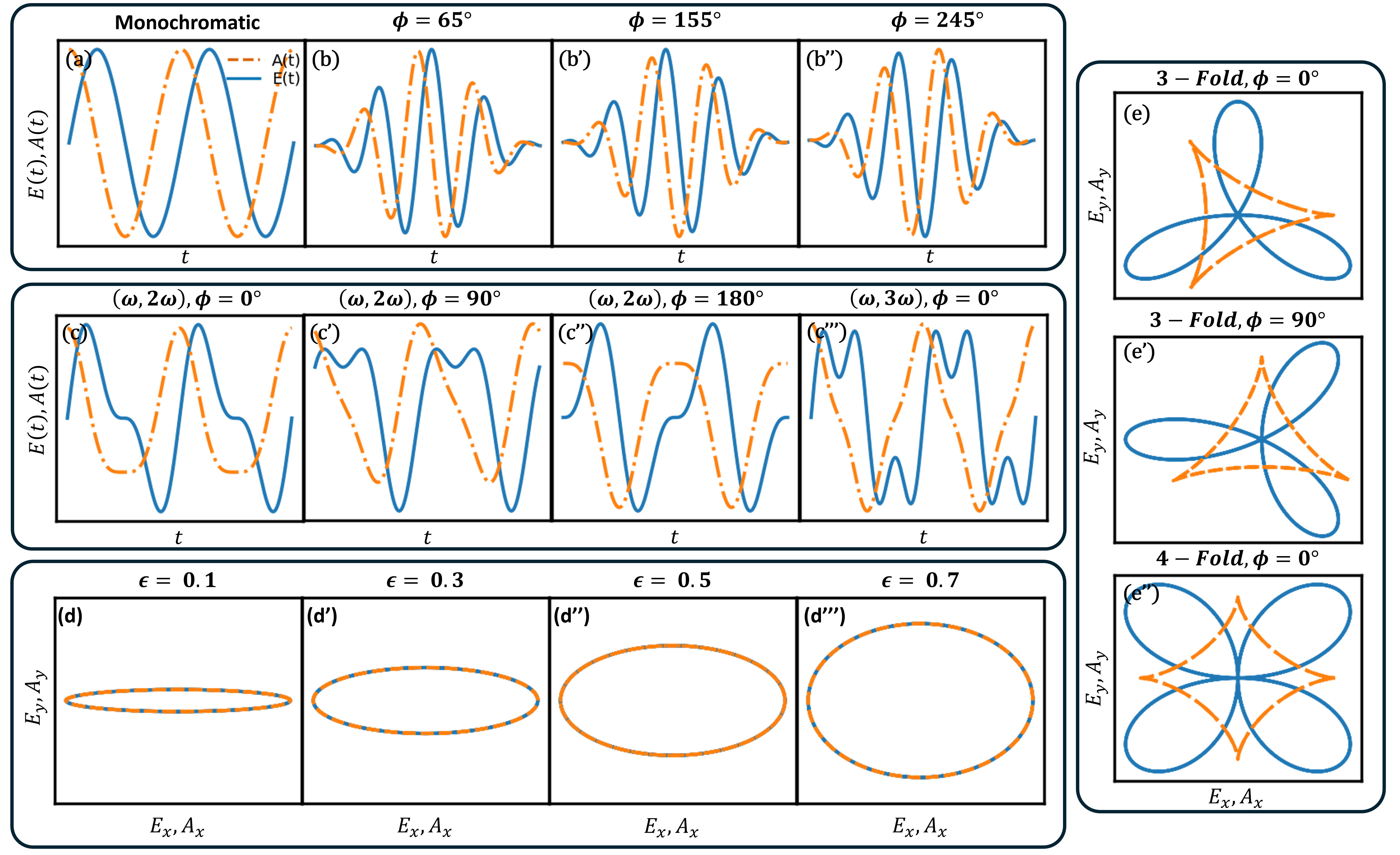}
    \caption{Electric field (blue solid line) and the corresponding vector potential (orange dashed line as functions of time for the 
    (a) monochromatic field, (b-b'') few-cycle pulse with $N=4.3$ and carrier-envelope phases $\phi = 65^{\circ}, 155^{\circ}$ and $245^{\circ}$, (c-c''') ($\omega, 2\omega$) and $(\omega, 3\omega)$ bichromatic linearly polarized fields with ratio of field amplitudes $\xi = 0.8$ and relative phase $\phi$ as indicated in the panels, (d-d''') elliptically polarized fields with ellipticity $\varepsilon = 0.1, 0.3, 0.5$ and $0.7$  and (e-e'') counter-rotating bicircular fields with identical amplitude, relative phase $\phi$ and commensurate frequency $s$ as indicated in the panels. For all fields, we use wavelength $\lambda = 800$nm ($\omega = 0.057$) a.u., and (peak) intensity $I = 1.0 \times 10^{14}$ W/cm$^2$.  }
   \label{fig:fields_overview}
\end{figure*}

In this section, we introduce the general symmetry operators and discuss how the symmetries manifest in the electric field profile in time, of each of the tailored fields with which direct ATI is investigated in this work. Although these refer to geometric properties of the waveform, they translate directly into dynamical symmetries of the laser-driven system since the field determines the time dependence of the equations of motion. Consequently, this informs our later discussion of the direct ATI ionization times computed using the PINN solver, and the corresponding PMDs. 

\begin{enumerate}
    \item The\textbf{ time-translation} operator $\hat{\mathcal{T}}_T(\tau_T)$ is defined as 
\begin{equation}
\hat{\mathcal{T}}_T(\tau_T): \quad t \rightarrow t + \tau_T ,
\label{eq:timetranslation}
\end{equation}
which shifts the electric field waveform along the time axis, $E(t) \rightarrow E(t+\tau_T)$. If the driving field is invariant under $\hat{\mathcal{T}}_T(\tau)$, the equations of motion are unchanged by this transformation, and solutions occur in symmetry-related `families' separated by the time shift $\tau$. 
\item The\textbf{ time-reflection} operator $\hat{\mathcal{T}}_R(\tau_R)$ is defined as
\begin{equation}
\hat{\mathcal{T}}_R(\tau_R): \quad t \rightarrow -t + \tau_R ,
\end{equation}
which reflects the electric field waveform about a reference time $\tau_R$, $E(t) \rightarrow E(-t+\tau_R)$. Time-reflection symmetry relates ionization and recollision events occurring before and  after the reference time. 
\item For linearly polarized fields, the \textbf{field-inversion} operator $\hat{\mathcal{F}}$ is defined as 
\begin{equation}
\hat{\mathcal{F}}: \quad E(t) \rightarrow -E(t), A(t) \rightarrow - A(t)
\end{equation}
which is an inversion of the electric field (or vector potential) along the polarization axis. In the dipole approximation and for inversion-symmetric targets, this operator corresponds to a reflection about the time-axis. It is dynamically equivalent to inversion of momentum along the polarization direction.
\item For nonlinearly polarized fields, the discrete rotation operator $\hat{R}_n$ plays a role, and is defined as,
\begin{equation}
\hat{R}_n: \quad \mathbf{E}(t) \rightarrow \mathcal{R}_{\theta_n}\mathbf{E}(t) ,
\end{equation}
where $\mathcal{R}_{\theta_n}$ denotes a rotation by an angle $\theta_n$. Note that while $\hat{R}_\theta$ is a generic rotation operator, rotational symmetry only exists for specific angles $\theta = \theta_n$. For linearly polarized fields, rotating by any angle changes the field. Similarly, for elliptically polarized fields, rotations change the ellipse orientation and so there is no discrete rotational symmetry in general. However, for bicircular fields,  this can exist and will be elaborated upon below. 
\end{enumerate}

All physically relevant dynamical symmetries of the driving fields considered in this work can be constructed from combinations of the operators defined above. These are:
\begin{enumerate}
    \item \textbf{Half-cycle symmetry}. This is translation of half-cycle followed by a reflection along the time-axis $\hat{\mathcal{F}}\,\hat{\mathcal{T}}_T(T/2)$. This can also be expressed as $E(t\pm T/2) = -E(t)$. 
\item \textbf{Time-reflection about extrema}. This is defined by invariance of the driving field at $\hat{\mathcal{T}}_R (\tau_{ex})$ where $\tau_{ex}$ are the times where the extrema occur.
\item \textbf{Time-reflection about zero crossings, followed by reflection along the time-axis,} $\hat{\mathcal{F}}\hat{\mathcal{T}}_R(\tau_{cr})$ where $\tau_{cr}$ are the times of the zero crossings.
\item \textbf{Rotation-time-translation.} This is time-translation followed by a discrete rotation in the polarization plane $\hat R_n\,\hat{\mathcal T}_T(T/n)$. If the field is invariant under this operator then the system exhibits a discrete rotational symmetry of order $n$.
\end{enumerate} 
Note that these are only a subset of the possible operators and combinations - see \cite{Neufeld2025} for a more thorough catalogue. 

More generally, tailored fields may preserve or break individual elementary symmetries while retaining specific combined operations. Identifying which operators remain valid for a given field configuration provides a systematic way to anticipate the electron dynamics, that is, which saddle point solutions are identical, which events are dominant or suppressed, and hence provides a useful way to simplify computations for more complex fields, but also aids prediction of which trajectories will interfere and what the resulting patterns may be. 
We shall now discuss the symmetries of selected fields more thoroughly. These are displayed in Fig.~\ref{fig:fields_overview}.

\subsection{Linearly polarized fields}
\label{subsec:linear_fields}
\subsubsection{Monochromatic fields}
\label{subsec:monotailored}

Linearly polarized monochromatic fields provide a useful reference point for the upcoming discussion.  These fields take the form,
\begin{equation}
    \textbf{E}(t) = E_0 \sin(\omega t)\,\mathbf{\hat{e}}_x.
    \label{eq:monofield}
\end{equation}
where $E_0$ is the field amplitude, $\omega$ is the frequency and $\mathbf{\hat{e}}_x$ denotes the polarization direction. Owing to their strict periodicity and simple temporal structure, these fields exhibit three key symmetries: (i) half-cycle symmetry $E(t\pm T/2)=-E(t)$, (ii) symmetry regarding a time reflection around its extrema, so that $\hat{\mathcal{T}}_R\left(\tau_{ex}\right)E(t)=E(t)$,  and (iii) invariance with respect to a time reflection around its zero crossings followed by a reflection with regard to the time axis $\hat{\mathcal{F}}\hat{\mathcal{T}}_R\left(\tau_{cr}\right)E(t)=E(t)$. These symmetries imply equivalence between ionization and rescattering events separated by half cycles and lead to highly constrained momentum-space structures. They also hold for the vector potential $\textbf{A}(t)$ although, of course, the times will be different.

\subsubsection{Bichromatic fields}
\label{subsec:bichromatictailored}
Linearly polarized bichromatic fields take the form,
\begin{equation}
\textbf{E}(t) = [E_1 \cos(r\omega t) + E_2 \cos(s\omega t + \phi)] \mathbf{\hat{e}}_x,
\label{eq:bichromatic_field}
\end{equation}
where $r$ and $s$ denote the harmonic order of the first and second colors respectively and $\phi$ is the relative phase between the two components. Symmetries can be preserved or broken by tuning $\phi$ and changing the temporal structure of the field. This phase controls how different features of the constituent fields align in time, such as their zero crossings and extrema. Therefore, time-reflection symmetry only occurs at certain values of the relative phase. Half-cycle symmetry is only retained if the field satisfies $E(t\pm T+/2) = -E(t)$ which requires $r+s$ to be even.  In \cite{Rook2022}, this was shown to depend on the parity of $(r,s)$ with specific conditions for $\phi$ provided.  Here we focus on fields for which $r=1$ and $s=2$ or $3$. Consequently, $(\omega, 2 \omega)$ and $(\omega, 3\omega)$ fields exhibit qualitatively different symmetry behavior as the relative phase $\phi$ is varied. The symmetries for the specific bichromatic fields of interest in this work are summarized in Table~\ref{tab:bichromatic_symmetries}.

\begin{table}[t]
    \centering
    \begin{adjustbox}{max width=\columnwidth}
    \begin{tabular}{llll}
        \toprule\toprule
        Field & Crossings & Extrema & Half-cycle \\
        \midrule
        $(\omega, 2\omega, \phi=\pi)$        & \ding{51} & \ding{55} & \ding{55} \\
        $(\omega, 2\omega, \phi=\pi/2)$      & \ding{55} & \ding{51} & \ding{55} \\
        $(\omega, 2\omega, \phi=3\pi/4)$     & \ding{55} & \ding{55} & \ding{55} \\
        \midrule
        $(\omega, 3\omega, \phi=0)$          & \ding{51} & \ding{51} & \ding{51} \\
        $(\omega, 3\omega, \phi=\pi/2)$      & \ding{55} & \ding{55} & \ding{51} \\
        $(\omega, 3\omega, \phi=-\pi/2)$     & \ding{55} & \ding{55} & \ding{51} \\
        \bottomrule
    \end{tabular}
    \end{adjustbox}
    \caption[Dynamical symmetries of selected linearly polarized bichromatic fields]{Dynamical symmetries of selected linearly polarized bichromatic fields
for different values of the relative phase $\phi$ used in this work.}
\label{tab:bichromatic_symmetries}
\end{table}

\subsubsection{Few-cycle pulse}
\label{subsec:pulsetailored}
Few-cycle pulses can take the form,
\begin{equation}
\begin{aligned}
\textbf{E}(t) &= E_0 \big[\sin^2\big(\frac{\omega t}{2N}\big) \sin(\omega t + \phi)] \mathbf{\hat{e}}_x \\&= E_0 \big[ \sin(\omega t +\phi) - \frac{1}{2} [\sin (\omega t (1+ \frac{1}{N}) + \phi) \\&+\sin(\omega t (1-\frac{1}{N}) + \phi)\big] \mathbf{\hat{e}}_x
    \label{eq:pulseEquation}
\end{aligned}
\end{equation}
where $N$ is the number of optical cycles and $\phi$ is the carrier envelope phase.  Here, we use the convention in \cite{Hashim2024}, which sets $\phi=\phi_1-\phi_0$, with $\phi_0=60^{\circ}$. The few-cycle pulse can be decomposed into a linear combination of three monochromatic waves, each of which carry the CEP throughout. These fields differ significantly from the monochromatic and bichromatic cases, in that the field envelope breaks their periodicity. Therefore, different field cycles are not equivalent $E(t+T) \neq E(t)$.
They lose their time-translation symmetry, and with it the half-cycle symmetry. Time-reflection symmetry about the crossings and extrema are also generally broken, except in highly specific cases. such as when $\phi = n\pi$ where $n$ is an integer, for linearly polarized pulses. Time reflection symmetry will only occur if the field envelope is symmetric and the carrier is even under time reversal.

The implication of this symmetry breaking on the saddle point solutions is that ionization and rescattering events occurring in different cycles of the pulse are no longer equivalent. The dynamics of the process are dominated by a small set of events associated with where the field is the strongest (as this influences where the ionization and/or return probabilities are the highest). Few-cycle pulses can therefore be said to change the `dominance' of different events. Because the degeneracy between events in different cycles is lifted relative to the monochromatic case, one can make statements about which cycle is dominant, which events are suppressed and can be neglected, and so on. Predictions on interference can also be made. If not all events contribute equally, then it is legitimate to consider just the dominant ones. Hence, it is easier to disentangle interference arising from different events, and in particular to disentangle intra- and inter-cycle interference.

\subsection{Elliptically polarized fields}
\label{subsec:ellipticaltailored}
Elliptically polarized fields introduce a transverse component of the electric field,
\begin{equation}
    \mathbf{E}_{\varepsilon}(t) = E_0 \left[ \sin(\omega t)\,\hat{\mathbf{e}}_x + \varepsilon \cos(\omega t)\,\hat{\mathbf{e}}_y \right].
    \label{eq:elliptical}
\end{equation}
where $\varepsilon \in [0,1]$ is the field ellipticity and $\hat{\textbf{e}}_x$ and $\hat{\textbf{e}}_y$ are the unit vectors along the major and minor polarization axes respectively. This formulation ensures that for $\varepsilon=0$ we recover a linearly polarized monochromatic field, while $\varepsilon=1$ represents circular polarization and has perfect continuous rotational symmetry which breaks with ellipticity. Half-cycle symmetry is retained regardless of ellipticity.

Elliptically polarized fields find use in controlling recollision probability, and introduce transverse drift momentum, possibly altering the transverse momentum structure in one-electron distributions. While few-cycle pulses primarily change the dominance of events, elliptically polarized fields can change whether recollision happens at all.

\subsection{Bicircular fields}
\label{subsec:bicirculartailored}
We now take a brief foray into fields exhibiting discrete rotational symmetry, namely bicircular fields. These are formed by the superposition of two circularly polarized components with commensurate frequencies $r$ and $s$,
\begin{equation}
\begin{aligned}
   \textbf{ E}(t) &=&  E_1 [\cos(\omega t)\hat{\mathbf{e}}_x + \sigma_1\sin(\omega t)\hat{\mathbf{e}}_y] \\&+& E_2[ \cos(s \omega t + \phi) \hat{\mathbf{e}}_x + \sigma_2\sin(s\omega t + \phi)\hat{\mathbf{e}}_y]
\end{aligned}
\label{eq:bircular}
\end{equation}
where we have taken $r=1$, $\sigma_{1,2} = \pm1$ is the helicity (left or right handed rotation), and $\sigma_1 = - \sigma_2$ denotes counter-rotating fields. Here $E_1$ and $E_2$ are the amplitudes of the two colors, the relative phase $\phi$ defines the alignment in time of the two colors similar to the bichromatic case.

By changing $s$ and $\phi$, one can tune these types of fields such that they remain invariant under time translation by $T/n$ and rotation of the polarization plane by $2\pi/n$ where $n=(s+1)$ \cite{Orth2021, Madsen2016}.  Therefore, the field exhibits a discrete rotational symmetry of order $n=s+1$,
\begin{equation}
\begin{aligned}
\hat R_n\,\hat{\mathcal T}_T(T/n)\textbf{E}(t) &= E_1 \begin{pmatrix}
    \cos(\omega t) \\ \sin(\omega t)
\end{pmatrix} \\&+  E_2 \begin{pmatrix}
     \cos(n\omega t + \phi - 2\pi/n) \\ -\sin(n\omega t + \phi - 2\pi/n)
\end{pmatrix}
    \label{eq:bicircularInvariance}
\end{aligned}
\end{equation}
$\hat R_n\,\hat{\mathcal T}_T(T/n)\textbf{E}(t)  = \textbf{E}(t)$ if and only if $\phi \equiv \phi - 2\pi/n$ (mod $2\pi$) which is not satisfied for arbitrary $\phi$. 
This can be understood intuitively from the fact that the $\omega$ component rotates once per period while the $s\omega$ component rotates $s$ times per period, in the opposite direction. When the time translation and rotation is performed, for the discrete rotational symmetry to be retained, the total relative phase must be equal to $2\pi$. Note that the $(\omega, 2\omega, \phi=90^{\circ})$ bicircular field does not have true three-fold rotational symmetry as it does not satisfy the above constraint on $\phi$. Nonetheless, it can have time-reflection symmetry about the field extremum, meaning saddle point solutions may be linked by reflection symmetry but are not rotationally degenerate \cite{Habibovic2021}. Bicircular fields allow access to different symmetries and the consequence of their symmetry breaking on the electron dynamics and on ATI photoelectron momentum  distributions (PMDs) can be explored.

\section{Methodology}\label{sec:method}

\begin{figure*}[t]
    \centering
    \includegraphics[width=0.75\textwidth]{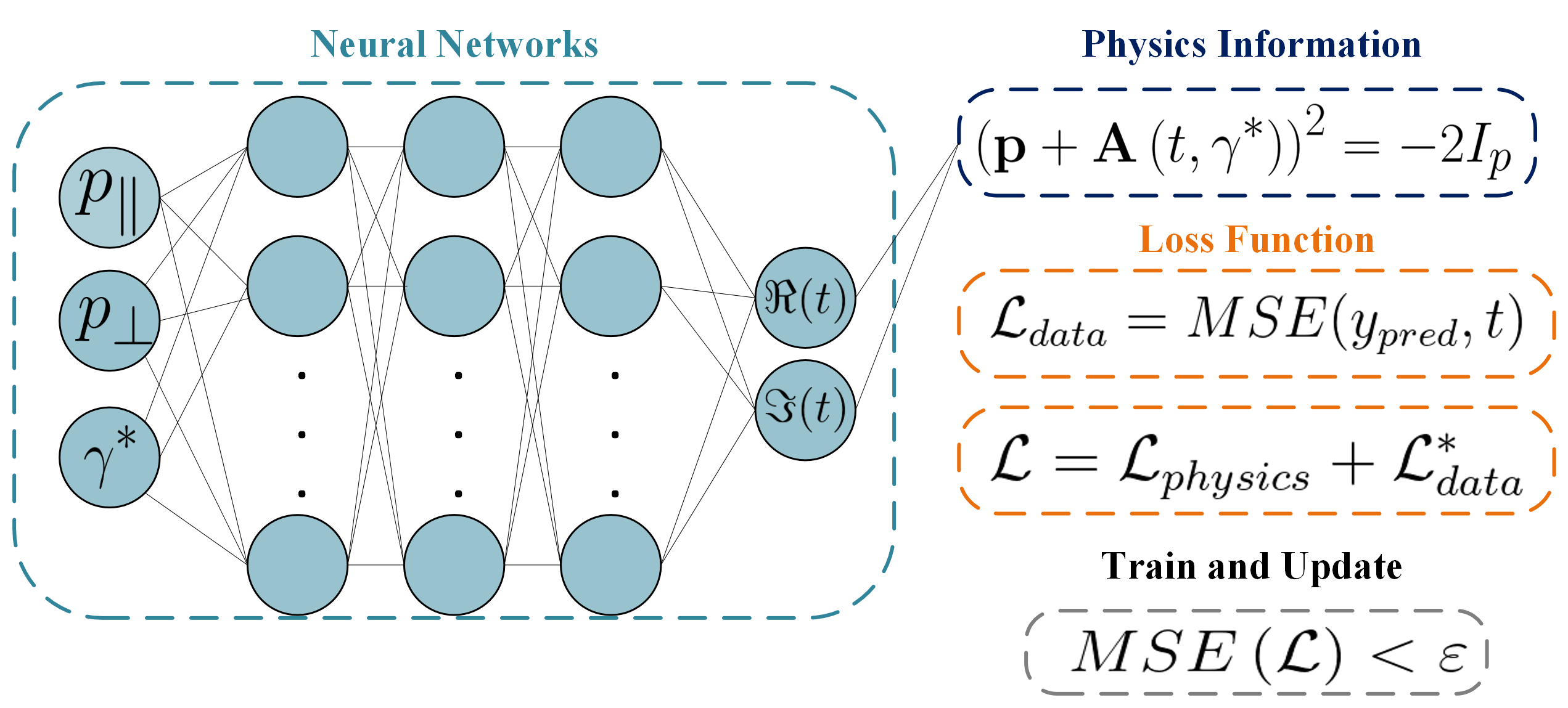}
    \caption[Schematic of the physics-informed neural network workflow]{Schematic of the physics-informed neural network with 3 hidden layers. Inputs comprise of the momentum components ($p_\parallel,p_\perp$) and field parameters $\gamma$. The network outputs the real and imaginary parts of the complex saddle time, $(t_{\mathrm{R}},t_{\mathrm{I}})$. Training minimizes a physics residual $\mathcal{L}_{\text{physics}}$ derived from the saddle-point equation until the mean square error falls below a threshold accuracy; a supervised data term $\mathcal{L}_{\text{data}}$ can be included when labeled solutions are available, but is omitted here.}
    \label{fig:pinn_workflow}
\end{figure*}

Reference solutions for the direct ATI ionization times are computed by solving Eq.~\eqref{eq:ati_spe} using a local gradient-based root-finding algorithm (Newton-type method). For such methods, practical performance depends strongly on the initialization. The initial guess-time for the solver is provided heuristically. The simple-man model provides approximate real ionization and return times in the limit of vanishing binding energy, which is subsequently deformed into the complex time plane using \textit{ad hoc} methods, to obtain physically meaningful saddle-point solutions. For relatively simple fields, such as linearly polarized fields, physically relevant saddles can often be followed across intensity or other parameters via extrapolation. Convergence can be aided by using solutions from previous parameter-space grid points. For tailored fields and multi-parameter scans, this approach becomes substantially less reliable. 
Additionally, these methods are likely to break down when there are many nearby solutions or if the solutions are rapidly-changing. This is particularly problematic for HHG or ATI with rescattering, which are not discussed here.  The SFA SPEs for strong field processes are nonlinear and do not necessarily satisfy the conditions for which numerical convergence using Newton-type methods is guaranteed. This motivates the need to seek an alternative approach to solving the SPEs.

\subsection{Physics-informed neural networks}\label{subsec:pinn_background}

A physics-informed neural network is a NN architecture that incorporates physics information directly into its training process \cite{Raissi2019,Cuomo2022}. PINNs, like ordinary NNs can be trained on data, as well as governing physical equations of the system which are typically ordinary or partial differential equations or conservation laws. Instead of learning purely from the labeled data, the network is penalized when it violates the physics. The central idea of a PINN is to reformulate the solution of a physical problem as an optimization task, in which a neural network is trained to output functions that satisfy the governing equations to some prescribed tolerance \cite{Raissi2019}.

The governing constraint here is the direct ATI saddle-point equation [Eq.~\eqref{eq:ati_spe}],
\begin{equation}
[\mathbf{p} + \mathbf{A}(t,\boldsymbol{\gamma})]^2 + 2 I_p = 0,
\label{eq:ati_spe_pinn}
\end{equation}
which relates the final electron momentum $\mathbf{p}$, the laser vector potential $\mathbf{A}(t,\boldsymbol{\gamma})$, parametrized by a set of field parameters $\boldsymbol{\gamma}$, and the ionization potential $I_p$ of the target. The solution of Eq.~\eqref{eq:ati_spe_pinn} yields the complex ionization time $t$ associated with a given final momentum and field configuration. The complex saddle time is written as
\begin{equation}
t = t_{\mathrm{R}} + i\, t_{\mathrm{I}},
\end{equation}
with $t_{\mathrm{R}}$ and $t_{\mathrm{I}}$ denoting its real and imaginary parts, respectively. Now, the PINN is trained to approximate the mapping
\begin{equation}
(\mathbf{p}, \boldsymbol{\gamma}) \;\longmapsto\; (t_{\mathrm{R}}, t_{\mathrm{I}}),
\end{equation}
where the inputs consist of the final electron momentum components and the relevant laser-field parameters. A schematic of the network architecture and training loop is shown in Fig.~\ref{fig:pinn_workflow}.

The training objective is constructed by penalising violations of Eq.~\eqref{eq:ati_spe_pinn} through a physics residual. In the most general form, we consider a composite objective
\begin{equation}
    \mathcal{L} = \mathcal{L}_{\text{data}} + \lambda\,\mathcal{L}_{\text{physics}},
\end{equation}
where $\mathcal{L}_{\text{data}}$ is an mean square error (MSE) term defined when labeled saddle times are available, and $\mathcal{L}_{\text{physics}}$ enforces the constraint
\begin{equation}
    \mathcal{L}_{\text{physics}} = \left\| (\mathbf{p} + \mathbf{A}(t_{\text{pred}}, \gamma))^2 + 2I_p \right\|^2.
\end{equation}
where $t_{\mathrm{pred}} = t_{\mathrm{R}} + i t_{\mathrm{I}}$ denotes the complex time predicted by the network. Minimizing $\mathcal{L}_{\mathrm{physics}}$ therefore drives the network output towards solutions of  Eq.~\eqref{eq:ati_spe_pinn}. 

In this work, we focus on the unsupervised configuration by setting $\mathcal{L}_{\text{data}}=0$ and training using $\mathcal{L}_{\text{physics}}$ alone. The network is optimized via backpropagation with gradient-based methods, using automatic differentiation to propagate gradients through the residual. Training is terminated once the loss decreases below a prescribed tolerance $\varepsilon$. This choice removes the requirement to generate large training datasets with classical solvers, and it is naturally suited to testing generalization across different driving-field shapes.

We define the learning task as approximating a mapping $f: \mathbb{R}^{2+n} \to \mathbb{R}^2$ \footnote{Here $n=3$ as we take only the laser intensity but it can be any number, in principle.}. The input vector $\mathbf{x}$ comprises various laser field parameters such as the (logarithmic) intensity, field carrier envelope phase, ellipticity, relative phase and the momentum grid coordinates,
\begin{equation}
    \mathbf{x} = [\,\gamma,\, p_\parallel,\, p_\perp\,]^\top.
\end{equation}
The target output $\mathbf{y}$ represents the complex time response $t$, decomposed into real and imaginary components to facilitate learning with real-valued neural networks,
\begin{equation}
\mathbf{y} = [\, \text{Re}(t),\, \text{Im}(t) \,]^\top.
\end{equation}
Prior to training, inputs are standardized. The resulting feature vectors are normalized to ensure numerical stability.

\subsection{Implementation Details and Model Optimization}
\label{sec:impl}

Here, we detail the design choices that enable stable learning of multiple saddle-point solutions and present the ablation study used to select the final network configuration. A central practical difficulty is that the mapping from inputs (momenta and field parameters) to saddle points times is multi-valued: for a given input, several physically valid complex roots may exist, and a naïve regressor may collapse onto a single root. Our implementation therefore incorporates an explicit mechanism to control which solution root the network is trained to represent.

Existing Newton-type solvers depend heavily on the initial guess, with different initial conditions converging to different roots. Saddle point solutions are also strongly dependent on the driving field, for instance, for a monochromatic field, there exist two distinct ionization events per optical cycle, one in each half-cycle which are expected to be identical up to a phase. For more complex tailored fields (e.g.\ bichromatic, as illustrated in Fig.~\ref{fig:fields_overview}(c)), more than two roots per optical cycle may appear.  A PINN formulation trained only through the saddle point residual i.e. without any supervision, labels, or symmetry constraints, the network exhibits mode collapse, that is, it learns just one solution and ignores all other valid solutions. The PINN behaves as a global optimizer, and is naturally biased to roots with large smooth basins of attraction. Thus, in this case, it preferentially predicts the root associated with a \textit{dominant} basin of attraction. This is commonly a saddle with the smallest imaginary (or real) part as these are easiest for the network to learn. Other physically relevant solutions may be neglected.

\begin{figure}
    \centering
    \includegraphics[width=1\linewidth]{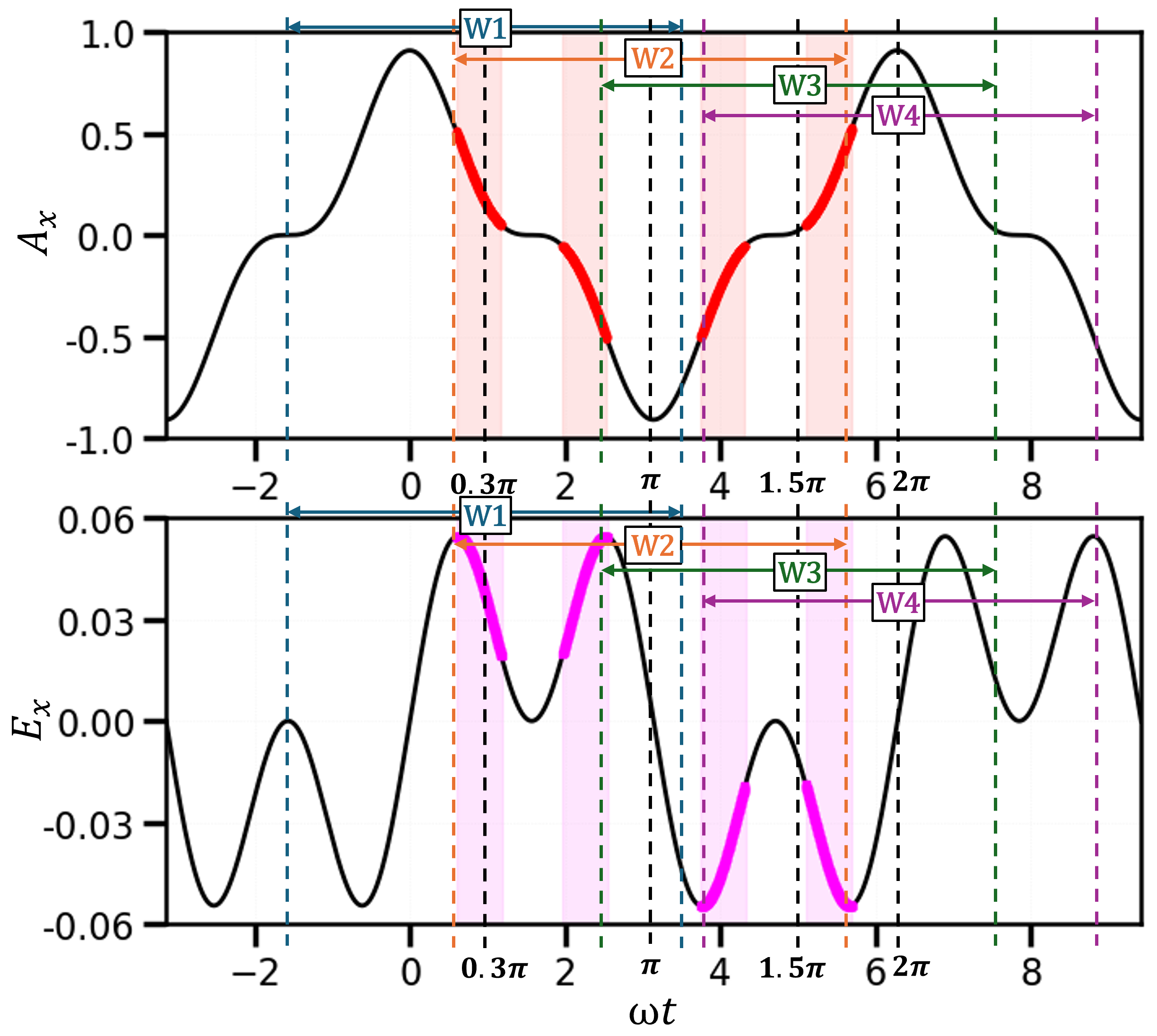}
    \caption{Window configuration for the \(x\)-direction saddle-time solutions in a four-fold bicircular field. The scatter points represent the complex-time solutions of the saddle-point equation. The four windows, labeled W1--W4, define the prescribed output regions used in the windowed parameterization. For each window, the black dashed vertical line marks the window center, while the colored dashed lines and arrows indicate the corresponding window range. The upper and lower panels show the two sets of solution branches in the \(x\)-direction.}
    \label{fig:window}
\end{figure}

To address this, we introduce a windowed parameterization that acts as a root selector. Rather than outputting the saddle time directly, the network predicts an unconstrained latent variable which is mapped into a prescribed region of the complex-time plane:
\begin{equation}
t_{\mathrm{pred}} \;=\; t_{\mathrm{c}} \;+\; s \odot g\!\left(\tilde{t}\right),
\end{equation}
where $t_{\mathrm{c}}$ denotes the window center, $s$ the window scale, $\tilde{t}$ the raw network output, $g(\cdot)$ a smooth squashing function (e.g.\ component-wise $\tanh$), and $\odot$ denotes element-wise multiplication on $(t_\mathrm{R},t_\mathrm{I})$. In Fig~\ref{fig:window}, we show the window configurations for $x$ component of four-fold bicircular field. Intuitively, $t_{\mathrm{c}}$ determines the cycle (or half-cycle) region explored by the model, while $s$ constrains the admissible deviation around that center. This parameterization provides two benefits: (i) it reduces ambiguity by restricting optimization to a single root manifold, and (ii) it improves conditioning by preventing the network from exploring large, highly nonlinear regions of the residual landscape.
In contrast to existing methods, this approach does not require precise initial guesses: coarse priors such as centers near $\pi/\omega$ or $\pi/(2\omega)$ are sufficient in practice. The choice of $t_{\mathrm{c}}$ and $s$ is guided by field symmetries (e.g.\ half-cycle symmetry, or symmetry about field extrema), which determine the expected structure and periodicity of the saddle points roots.

\subsection{Architecture and Hyperparameter Selection via Ablation}
\label{sec:ablation}
We perform an ablation study on the monochromatic field, where reference solutions are reliably obtained with existing solvers and the multi-solution structure is well understood. Our baseline is a fully connected network with $L=3$ hidden layers and width $W=128$ (per layer), trained without windowing. We then vary output constraints/activations, network capacity, and optimization hyperparameters. Table~\ref{tab:ablation} reports the final training loss and wall-clock runtime for representative configurations.

\begin{table}[t]
    \centering
    \caption{Ablation study for model selection in the monochromatic field. `Loss' denotes the final physics-residual objective under a fixed epoch budget. `Runtime' is the corresponding wall-clock training time measured on the same hardware. The windowed parameterization yields the largest improvement, indicating that explicit root control is critical in multi-root regimes.}
    \label{tab:ablation}
    \begin{tabular}{l c c}
        \toprule
        \textbf{Configuration} & \textbf{Loss} & \textbf{Runtime} \\
        \midrule
        Baseline model ($L=3, W=128$) & $3.42 \times 10^{-4}$ & 224s \\
        \midrule
        \textit{Constraints and activation} &  &  \\
        ReLU output & $4.12 \times 10^{-3}$ & 214s \\
        Softplus output & $2.80 \times 10^{-4}$ & 214s \\
        Windowed output & $3.26 \times 10^{-6}$ & 224s \\
        \midrule
        \textit{Architecture} &  &  \\
        Shallow depth ($L=2$) & $1.44 \times 10^{-3}$ & 228s \\
        Wide layers ($W=256$) & $3.14 \times 10^{-6}$ & 270s \\
        \midrule
        \textit{Hyperparameters} &  &  \\
        High learning rate ($10^{-3}$) & $7.21\times 10^{-6}$ & 234s \\
        Batch size 256 & $8.62 \times 10^{-7}$ & 640s \\
        \bottomrule
    \end{tabular}
\end{table}

Output constraints dominate performance: the windowed output reduces the loss by approximately two orders of magnitude relative to the baseline at essentially unchanged runtime. This suggests that the primary failure mode is not limited by the size of network, but ambiguity in the target mapping. Without root control, optimization can reduce the residual by converging to a convenient basin rather than learning the intended root consistently.

 Architectural complexity behaves as expected. Reducing depth ($L=2$) degrades performance substantially, indicating under-parameterization for the nonlinear mapping. Increasing width ($W=256$) improves the loss to a level comparable with windowing but increases runtime, suggesting diminishing returns when root ambiguity is not addressed.

Optimization introduces practical trade-offs. A larger learning rate improves convergence under a fixed epoch budget, while smaller batch sizes can achieve lower loss but incur significantly longer runtimes. Since our downstream goal is high-throughput evaluation across large parameter scans, we adopt the windowed configuration as a favorable balance between accuracy and computational cost.

\subsection{Training Dynamics and Failure Modes}
\label{sec:dynamics}
\begin{figure}[t]
    \centering
    \includegraphics[width=0.49\textwidth]{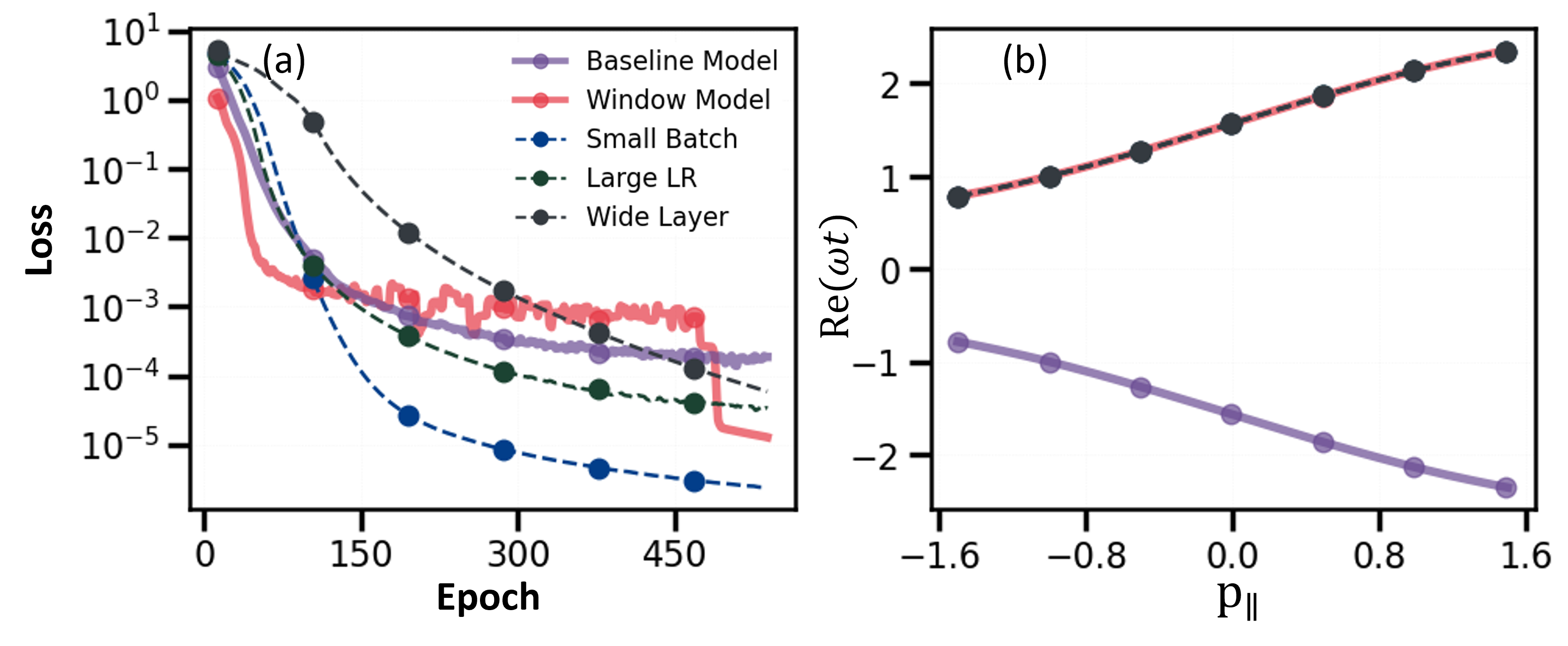}
    \caption[Training loss versus epoch and real part of the predicted ionization times using the PINN solver for direct ATI driven by monochromatic fields]{Training loss versus epoch for the five best-performing configurations of the PINN in Table~\ref{tab:ablation} [panel (a)] showing the model converges faster in early training and the real saddle point times $t_{\mathrm{R}}$ predicted by the PINN along $(p_\parallel, p_\perp=0)$ [panel (b)], showing that windowing enables consistent recovery of the desired root rather than collapsing to the nearest solution.}
    \label{fig:lossCurves}
\end{figure}

Figure~\ref{fig:lossCurves}(a) presents the loss curves for the baseline model, the model incorporating the window parametrization, as well as models using small-batch, high learning rate and wide layers, while Figure~\ref{fig:lossCurves}(b) shows the predicted real part $t_{\mathrm{R}}$ against the final parallel momentum of the electron ($p_\perp =0$) for the PINN solver and the existing Newton-type method. The windowed model exhibits a rapid decrease in loss within the first $\sim 100$ epochs and, importantly, is the only configuration that consistently converges to the intended non-trivial root rather than drifting toward a near-zero or incorrect solution.

Low residual loss alone does not guarantee physically meaningful outputs. Several non-windowed models reach losses in the range $10^{-4}$-$10^{-7}$ yet yield inconsistent predictions, including unstable root selection across intensities and sign flips in $t_{\mathrm{R}}$ across random seeds. This reflects the non-identifiability of the multi-root mapping when the objective provides no explicit mechanism to distinguish solution roots. Windowing mitigates this issue by restricting Optimization to a local neighbourhood in which the target root is represented uniquely.

Based on Table~\ref{tab:ablation} and the diagnostics in Fig.~\ref{fig:lossCurves}, we adopt the windowed-output model with $L=3$ and $W=128$ as the default configuration for the subsequent study. This choice provides robust root control, stable convergence behavior, and a practical runtime suitable for repeated evaluations in multi-parameter scans.

\subsection{Scaling and Training Complexity}
\label{sec:scaling}
We next examine how field complexity and input dimensionality affect training cost. More complex tailored driving fields can, in principle, increase optimization difficulty by introducing additional saddle roots per optical cycle, stronger nonlinearity in the saddle-point residual, and sharper parameter-dependent transitions between basins of attraction. A priori, these effects could slow training substantially and reduce robustness.

\begin{table}[t]
    \centering
    \caption{The number of inputs (conditioning variables provided to the network) and runtime (in seconds, measured on the same hardware under a fixed training budget  with the same optimizer and epoch count, and representing the wall-clock time to train a single root-specific model) for the fields and field parameters considered in Fig.~\ref{fig:fields_overview}.  Note that the runtime is the average runtime of all the specific field parameters for a given field type.}
    \label{tab:scaling_time}
    \begin{tabular}{l c c}
        \toprule
        \textbf{Field} & \textbf{Inputs $\gamma$} & \textbf{Runtime (s)}\\
        \midrule
        Monochromatic  & $[p_\parallel, p_\perp,I]$ & 224\\
        Few-cycle pulse & $[p_\parallel, p_\perp,I]$ & 210\\
        Bichromatic  & $[p_\parallel, p_\perp,I]$ & 217\\
        Elliptical & $[p_\parallel, p_\perp,I]$ & 210\\
        Bicircular  & $[p_\parallel, p_\perp,I]$ & 215\\
        \midrule
        Few-cycle pulse + CEP $ \phi$ & $[p_\parallel, p_\perp, I, \phi]$  & 1000\\
        Elliptical + $\varepsilon$ &  $[p_\parallel, p_\perp, I, \varepsilon]$  & 1500\\
        \bottomrule
    \end{tabular}
\end{table}
Table~\ref{tab:scaling_time} reports the average wall-clock time to train a single window across several field families, using the same optimizer, epoch budget and hardware\footnote{All experiments were conducted on a single NVIDIA GeForce RTX 3080 Laptop GPU.} For all cases with three inputs, namely $(p_\parallel, p_\perp, I)$ the runtime clusters around $210$-$224$s. This suggests that, once root selection is handled by the window constraints, the cost of learning a single saddle points root depends only weakly on the detailed driving field shape. For high-throughput scans, this is operationally useful: changing the field family does not substantially increase the training burden provided that the model is trained to represent one root and the window configuration is chosen appropriately.

We then extend the input vector to include additional continuous field parameters, enabling a single model to generalize beyond intensity scans. For few-cycle pulses, we include the carrier-envelope phase $\phi$ as a fourth input; for elliptically polarized fields, we include the ellipticity $\varepsilon$. This conditioning allows the trained network to predict saddle solutions for previously unseen combinations of intensity and the additional parameter, avoiding retraining separate models for each CEP or ellipticity setting. The trade-off is a substantial increase in training time: adding $\phi$ increases runtime from $\sim 210$s to $\sim 1000$s, and adding $\varepsilon$ increases runtime to $\sim 1500$s in our benchmarks [see Table~\ref{tab:scaling_time}]. This increase is consistent with the expanded input domain requiring denser sampling and slower convergence to maintain root-consistent predictions across the additional dimension.

Our benchmarks indicate that, under the proposed windowed-output constraints, training cost is largely insensitive to field complexity when the input dimensionality is fixed. The windowed-parameterization restricts the search region in complex time and enforces root consistency, stabilizing optimization across field families. As a consequence, for models trained with three inputs, the average runtime is comparable across all driving fields. This demonstrates the generalizability of the PINN solve and enables reuse of a single trained model for different tailored fields. The dominant factor is instead the number of distinct solution roots that must be covered: fields supporting more roots require more windows (or more specialized models) to recover all solutions, increasing total compute even if the per-window training time remains similar.

\section{Results}\label{sec:results}

In this section, we present the saddle point times and distributions for direct ATI obtained with the PINN solver discussed in Sec.~\ref{sec:method} for a range of tailored fields. These results are 
approached from the dual perspective of (i) demonstrating the robustness of the machine learning based solver and (ii) understanding the mapping of the field symmetries onto the ATI distributions. In Sec.~\ref{subsec:mllinear}, we focus on linearly-polarized fields whilst Sec.~\ref{subsec:mlnonlinear} discusses elliptically polarized and bicircular fields. 

We address key questions concerning the performance, robustness, and generalizability of the physics-informed neural network solver.
First, we examine how well the model learns the symmetries of the laser field. These symmetries are implicitly encoded through the window-output parametrization method, which identifies the extrema of the field corresponding to the real ionization times predicted by the simple-man model and searches for solutions within a surrounding window. The window width determines which half-cycles (or cycles) are accessible to the solver, and thus whether all physically relevant events can be identified. Second, for tailored fields such as few-cycle pulses, we investigate whether the model can correctly identify the dominant events as the CEP varies. This effectively tests whether the solver can detect changes in event dominance with relatively little guidance (i.e. with relatively little tuning of the window size), no additional physics constraints besides the field parameters, and without any labeled data (as it is the case here). Success would demonstrate robustness across a range of tailored fields and address the critical problem of generalizability that often limits conventional solvers.

Finally, we must answer how the model performs across varied laser field parameters. Here, we use a range of relative phases and harmonic frequencies for the bichromatic and bicircular fields, different CEPs for the pulse, different polarizations (ellipticities), as well as a range of intensities. This will enable us to answer whether the model is able to accurately predict the symmetries of the saddle point solutions in different parameter spaces, rather than being tuned to a specific field configuration.

Although the direct ATI saddle point solutions and distributions are well-established in the literature [see \cite{Eberly1991, Lewenstein1995, Ehlotzky2001, Milos2006, Bashkansky1987, Becker2017plane} for specific results of direct ATI with each of the tailored fields used here], we focus on them through the lens of symmetry, with the key question being that of how field symmetries map onto the saddle point solution structure and the ATI distributions. Through the elliptically polarized and bicircular fields, we introduce discrete rotational symmetries into the picture - how does this manifest in the ATI distributions?

A major advantage of our model is that, once trained for a given field configuration, it can predict saddle times for multiple laser field parameters such as intensity without retraining. 
The results have been compared against the `ground truth' (GT) computed using a verified Newton-type solver for DATI.

\subsection{Linearly polarized fields}\label{subsec:mllinear}
We now summarize the behavior of the saddle-point solutions and the resulting direct ATI momentum distributions within the strong-field approximation for linearly polarized driving fields. In the following, we display the times computed with $p_\perp=0$. However, the full momentum grid has been calculated.
\subsubsection{Monochromatic field}\label{subsubsec:results_mono}
\begin{figure*}[tph]
    \centering
    \includegraphics[width=\textwidth]{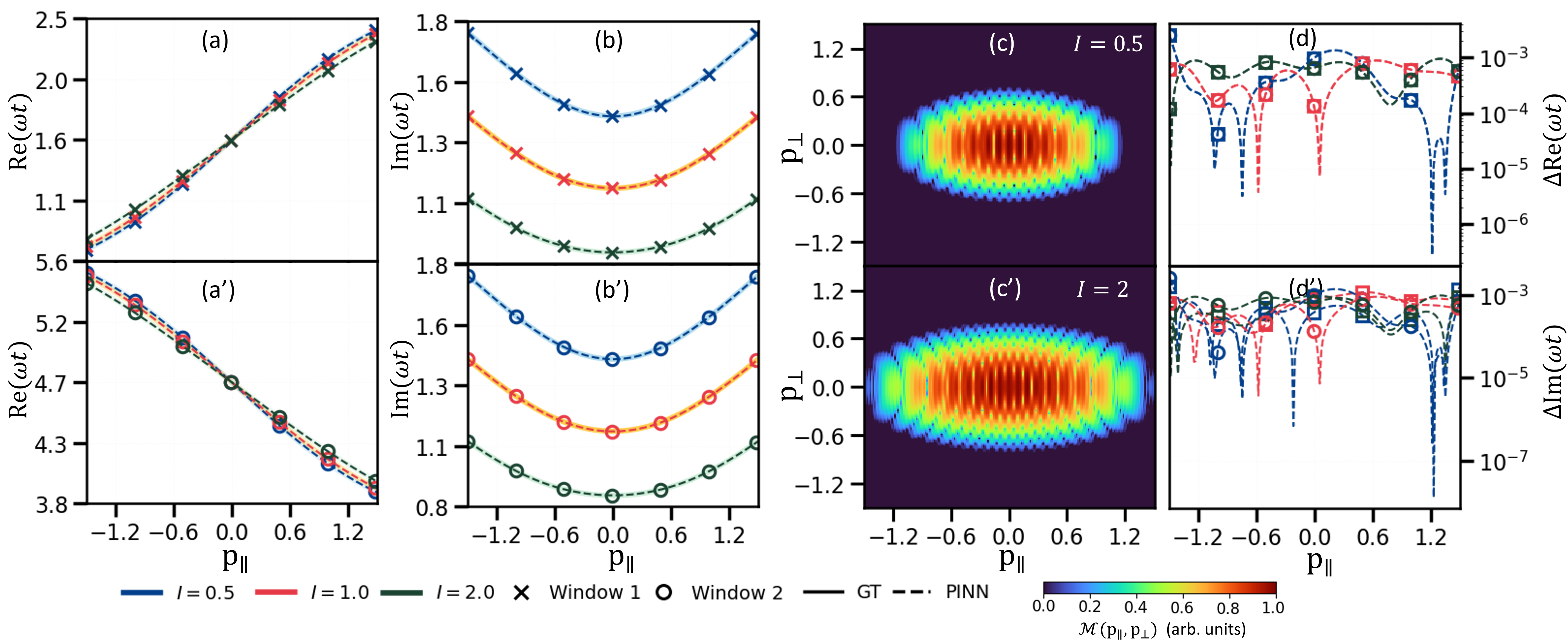}
\caption[Real and imaginary parts of ionization time, associated mean square error and photoelectron momentum distributions computed using the PINN solver for monochromatic fields]{The imaginary [panels (a),(a')] and real [panels (b), (b')] parts of the direct ATI ionization time computed at $(p_{\parallel}, 0)$ with the PINN solver detailed in Sec.~\ref{sec:method}, for two windows [first and second rows respectively, and distinguished using crosses and circles] in one cycle of the monochromatic field given by Eq.~\eqref{eq:monofield}. Three different intensities $I=0.5\times10^{14}$W/cm$^2$, $1\times10^{14}$W/cm$^2$ and $2 \times10^{14}$W/cm$^2$ have been used, denoted by blue, orange and black lines respectively. Panels (c), (c') show the PMD for the smallest and largest intensities respectively, while panels (d) and (d') display the mean square error in the real and imaginary parts of the time for all intensities, respectively. All other field parameters are the same as in the associated panel of Fig.~\ref{fig:fields_overview}.}
    \label{fig:mlmono}
\end{figure*}

Figs.~\ref{fig:mlmono}(a),(a') and (b),(b') show the real and imaginary parts of the ionization time as functions of the parallel electron momentum, respectively, predicted using the PINN solver detailed in Sec.~\ref{sec:method} for two half-cycles (and correspondingly, two windows) of the monochromatic field. 
The PINN predictions are in good agreement with the expected solutions computed using a standard root-finding method as the lines overlap exactly. Fig.~\ref{fig:mlmono}(d) and (d') shows the mean square error as a function of momentum, fluctuating between $10^{-3}-10^{-5}$, which is standard for PINN solvers \cite{Raissi2019} and does not cause any significant deviation of the predicted results from the expected. Panels (c) and (c') show the ATI photoelectron momentum distributions (PMDs) for intensities $0.5\times10^{14}$ and $2\times10^{14} \text{W/cm}^2$ respectively, computed using the PINN-predicted times for both half-cycles. The cutoff increases as the intensity increases, causing the distributions to be wider for larger intensities. This shows that the PINN solver has accurately learnt the structure of the saddle point times as function of the parameter-space (in this case the intensity-space). The ATI rings are visible as are the interference fringes, suggesting once again that the phase information (which arises from the saddle point times) is accurately captured. These results demonstrate that the solver yields physically meaningful observables across the full momentum grid and a range of intensities. Having established this as a baseline performance, we evaluate the robustness of the proposed solver.
This field satisfies the three symmetries $E(t+T/2)=-E(t)$, $\hat{\mathcal{T}}_R(\tau_{\mathrm{ex}})E(t)=E(t)$ and $\hat{\mathcal{F}}\hat{\mathcal{T}}_R(\tau_{\mathrm{cr}})E(t)=E(t)$ \cite{Rook2022}. As a result, there are two saddle point solutions per cycle, one in each half-cycle, which are each symmetric about $p_{\parallel}=0$. For this field, ionization happens very close to $\textbf{A}(t) \approx 0$ and so $\text{Im}[t]$ is smallest near $p_\parallel = 0$ and increases smoothly as $|p_{\parallel}|$ increases. The PMD is elongated along the polarization axis and is symmetric about $p_\perp =0$. To recover the correct symmetries in the ATI photoelectron momentum distributions, we perform unit‑cell averaging\footnote{Distributions are commonly averaged over a unit cell of the driving field. This is because experimentally measured quantities are insensitive to the absolute phase at which ionization occurs. Theoretical results explicitly depend on the choice of cycle or half-cycle through the saddle point solutions and accumulated phases, while experiments integrate over many ionization events at different phases. This procedure removes unphysical phase dependencies and allows more meaningful comparisons between theory and experiment \cite{Werby2021, Werby2022b}.} and propagate the field over several optical cycles. The resulting spectra exhibit well‑defined ATI rings arising from inter-cycle interference. As intensity increases, the cutoff also increases and the PMD becomes wider. All this behavior is as expected \cite{Becker2002Review}. 

\subsubsection{Few-cycle pulse with CEP}\label{subsubsec:results_cep}
\begin{figure*}
    \centering
    \includegraphics[width=\textwidth]{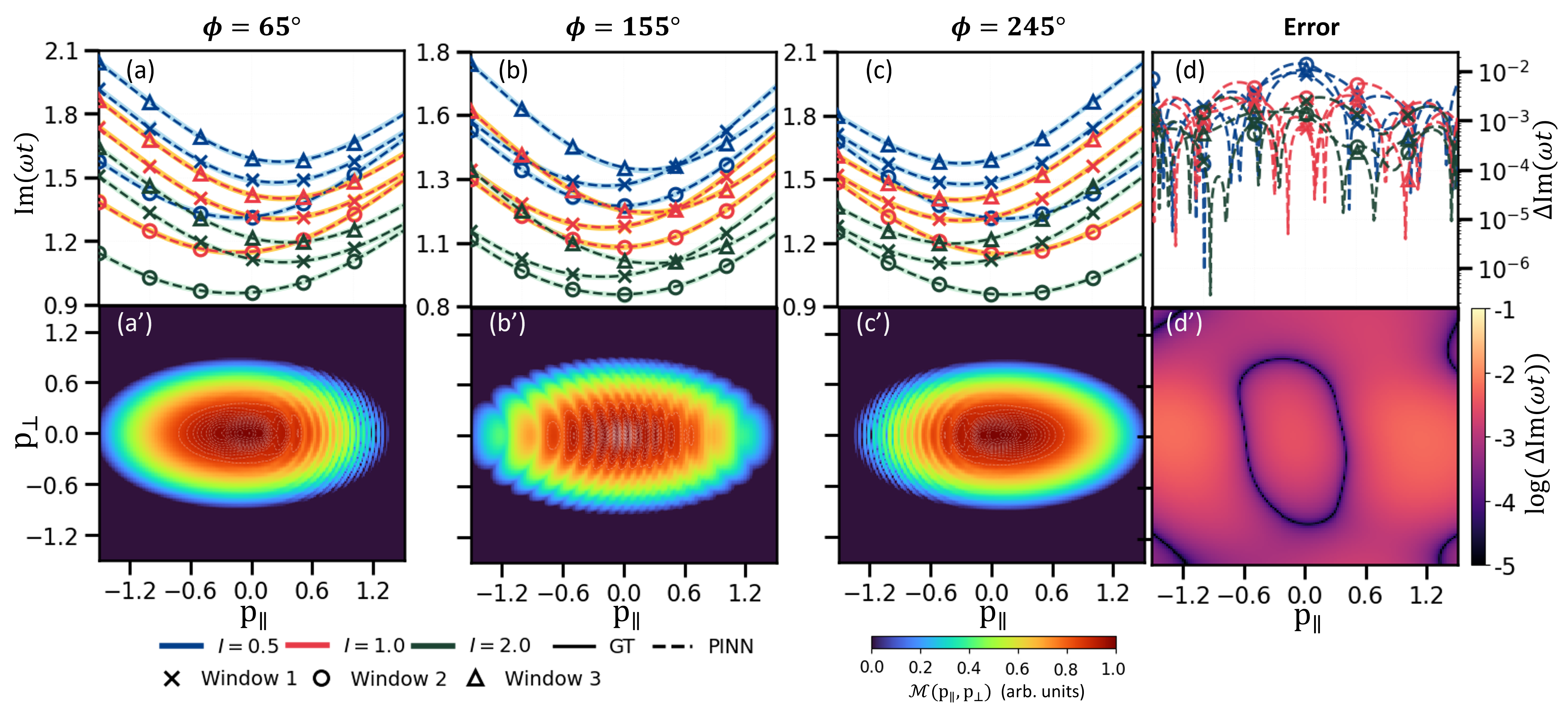}
    \caption[Imaginary parts of ionization time, associated mean square error and photoelectron momentum distributions computed using the PINN solver for the few-cycle pulse with varying CEPs]{The imaginary part of the direct ATI ionization time computed at $(p_{\parallel}, 0)$ with the PINN solver detailed in Sec.~\ref{sec:method} for the few-cycle pulses shown in Fig.~\ref{fig:fields_overview} [panels (a)-(c)]. For each field, multiple windows corresponding to each distinct solution are taken and denoted by crosses, circles and, triangles. Three different intensities $I=0.5\times10^{14}$W/cm$^2$, $1\times10^{14}$W/cm$^2$ and $2 \times10^{14}$W/cm$^2$ have been used, denoted by blue, orange and black lines respectively. The corresponding PMDs for each field are displayed in panels (a')-(c') with $I=2\times10^{14}$W/cm$^2$ and taking all windows and multiple cycles of the fields. Panels (d) and (d') show the mean square error (MSE) in the imaginary time at $(p_{\parallel}, 0)$ and for the entire 2D momentum grid with $\phi_1 =64^{\circ}$, respectively. All other field parameters are the same as in the associated panels of Fig.~\ref{fig:fields_overview}.}
    \label{fig:mlpulse}
\end{figure*}

The few-cycle pulse breaks all of the three dynamical symmetries present for monochromatic fields. Consequently, these saddle point solutions and distributions are mainly affected by the events taken into account.

Fig.~\ref{fig:mlpulse} exhibits the imaginary ionization times [panels (a)-(c)] and PMDs [panels (a')-(c')] for three CEPs of the few-cycle pulse. Pulses with $\phi_1=65^{\circ}$ and $245^{\circ}$ are reflections of each other in the time axis and hence their solutions and distributions are also equal and opposite. For each CEP, the three most dominant events have been found, using three windows. Events towards the center of the pulse, for which $\textbf{A}(t) \approx 0$ lead to imaginary times with minima at $p_\parallel\approx0$. For $\phi_1=155^{\circ}$, the vector potential is roughly equal and opposite for the events either side of the central peak, leading to a more symmetric PMD elongated along the polarization axis in which the interference between the events in one cycle are clearly visible, with the ATI rings playing a more secondary role. In contrast, for $\phi_1=65^{\circ}$ and $245^{\circ}$, the distributions have a clear asymmetry to the left or to the right, dependent on which events dominate. In these PMDs, the ATI rings are faintly visible.
The residual error plots shown in Figs.~\ref{fig:mlpulse}(d) and (d') confirm that the PINN solver consistently converges to physically meaningful saddle point solutions across all CEPs.
These results are in agreement with those for the second electron in \cite{Hashim2024}. The model is able to select the three (or however many windows are taken) most dominant events within the pulse and compute their ionization times. 

\subsubsection{Bichromatic fields}\label{subsubsec:results_bichromatic}

\begin{figure*}
    \centering
    \includegraphics[width=\textwidth]{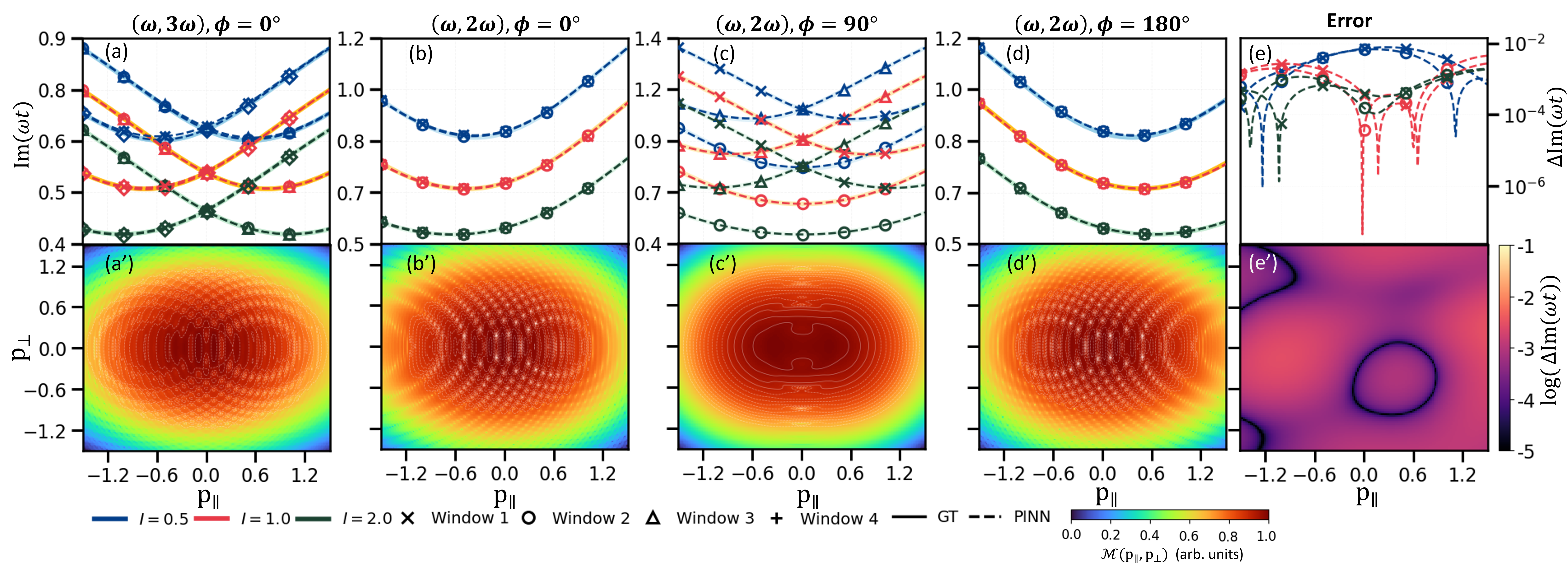}
   \caption[Imaginary parts of ionization time, associated mean square error and photoelectron momentum distributions computed using the PINN solver for bichromatic fields with varying harmonic frequencies and relatives phases]{The imaginary part of the direct ATI ionization time computed at $(p_{\parallel}, 0)$ with the PINN solver detailed in Sec.~\ref{sec:method} for the bichromatic fields shown in Fig.~\ref{fig:fields_overview}, [panels (a)-(d)]. For each field, multiple windows corresponding to each distinct solution are taken and denoted by crosses, circles, triangles and plus signs. Three different intensities $I=0.5\times10^{14}$W/cm$^2$, $1\times10^{14}$W/cm$^2$ and $2 \times10^{14}$W/cm$^2$ have been used, denoted by blue, orange and black lines respectively. The corresponding PMDs for each field are displayed in panels (a')-(d') with $I=2\times10^{14}$W/cm$^2$ and taking all windows and multiple cycles of the fields. Panels (e) and (e') show the mean square error (MSE) in the imaginary time at $(p_{\parallel}, 0)$ and for the entire 2D momentum grid with the $(\omega, 2\omega, \phi=2\pi)$ case, respectively. All other field parameters are the same as in the associated panels of Fig.~\ref{fig:fields_overview}.}
    \label{fig:mlbichromatic}
\end{figure*}

Figs.~\ref{fig:mlbichromatic}(a)-(d) shows the imaginary part of the ionization time computed with the PINN solver for the bichromatic fields shown in Fig.~\ref{fig:fields_overview}(c)-(c$'''$). For each field, multiple windows corresponding to each distinct solution \cite{Hashim2024b} are taken, denoted by the different symbols. Three different intensities $I=0.5\times10^{14}$W/cm$^2$, $1\times10^{14}$W/cm$^2$ and $2 \times10^{14}$W/cm$^2$ have been used, denoted by blue, orange and black lines respectively. The behavior of the times with intensity is as expected, and is the same as for the monochromatic case. This happens for all the tailored fields considered in the same way and will not be discussed further. Fig.~\ref{fig:mlbichromatic}(a')-(d') present the corresponding PMDs with the largest intensity, while the 1D and 2D mean square error (MSE) plots for the imaginary time are visualized in panels (e) and (e') for the $(\omega, 2\omega, \phi=2\pi)$ case as a sanity check. Since the error is very small at all values, the solver seems to have converged to the expected solution. 

\begin{table*}[t]
\centering
\renewcommand{\arraystretch}{1.25}
\setlength{\tabcolsep}{5pt}
\begin{tabular}{lllll}
\hline
\parbox[t]{2.6cm}{$(r,s,\phi)$} &
\parbox[t]{3.0cm}{$(1,3,0^\circ)$} &
\parbox[t]{3.0cm}{$(1,2,0^\circ)$} &
\parbox[t]{3.0cm}{$(1,2,90^\circ)$} &
\parbox[t]{3.0cm}{$(1,2,180^\circ)$} \\
\hline\hline
\parbox[t]{2.6cm}{Times} &
\parbox[t]{3.0cm}{$\times,\bigcirc:\; 0 \le t \le T/2$ \\ $\triangle,+:\; T/2 \le t \le T$} &
\parbox[t]{3.0cm}{$\times:\; 0 \le t \le T/2$ \\ $\bigcirc:\; T/2 \le t \le T$} &
\parbox[t]{3.0cm}{$\times,\bigcirc:\; 0 \le t \le T/2$ \\ $\triangle:\; T/2 \le t \le T$} &
\parbox[t]{3.0cm}{$\times:\; 0 \le t \le T/2$ \\ $\bigcirc:\; T/2 \le t \le T$} \\[2.5ex]
\hline
\parbox[t]{2.6cm}{$p_{\parallel}$} &
\parbox[t]{3.0cm}{$\times,\bigcirc:\; <0, >0$ \\ $\triangle,+:\; >0, <0$} &
\parbox[t]{3.0cm}{$\times:\; <0$ \\ $\bigcirc:\; >0$} &
\parbox[t]{3.0cm}{$\times,\bigcirc:\; >0,<0$ \\ $\triangle:\; 0$} &
\parbox[t]{3.0cm}{$\times:\; >0$ \\ $\bigcirc:\; >0$} \\[2.5ex]
\hline
\parbox[t]{2.6cm}{Events} &
\raisebox{-0.5\height}{\includegraphics[width=0.16\textwidth]{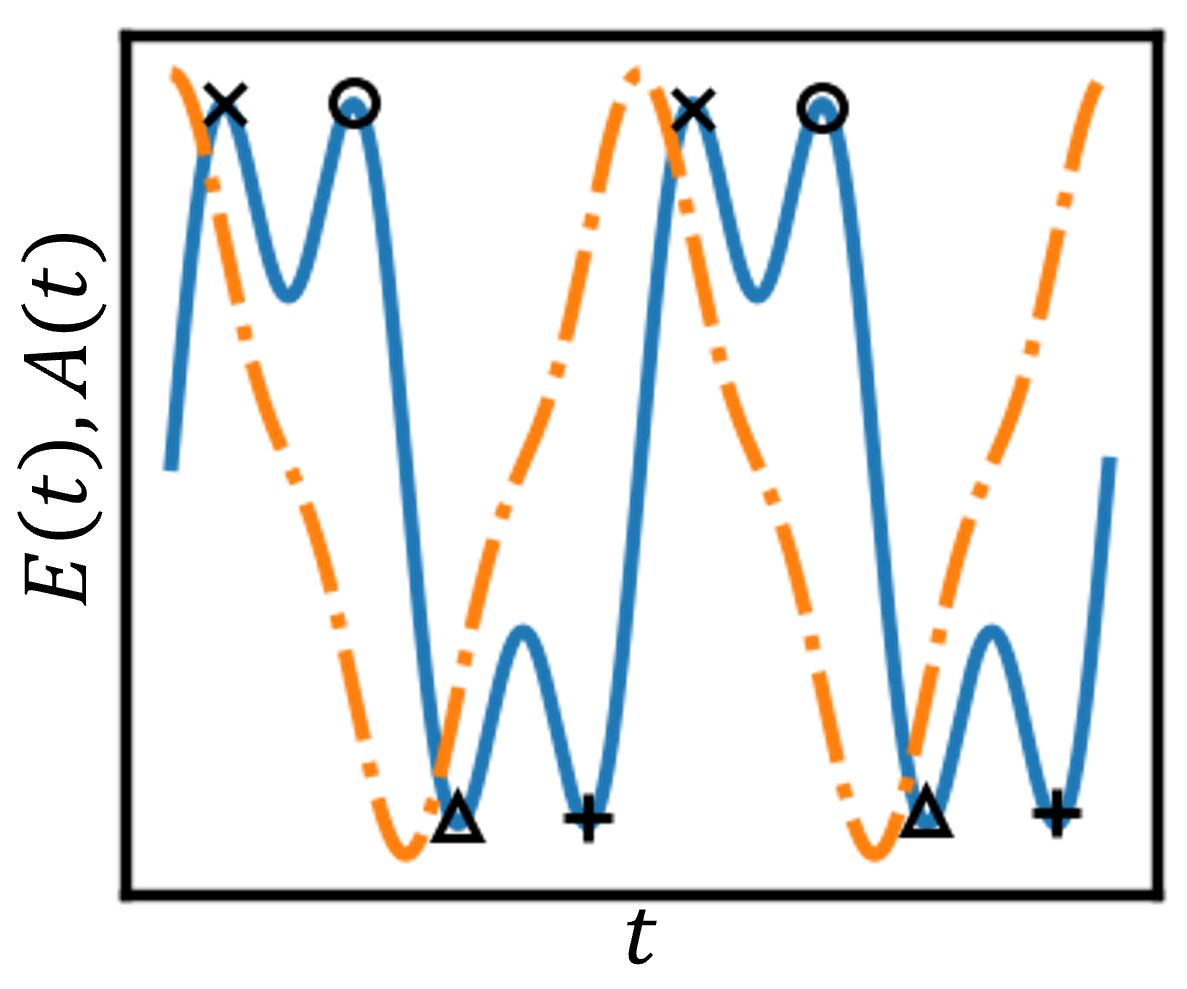}} &
\raisebox{-0.5\height}{\includegraphics[width=0.16\textwidth]{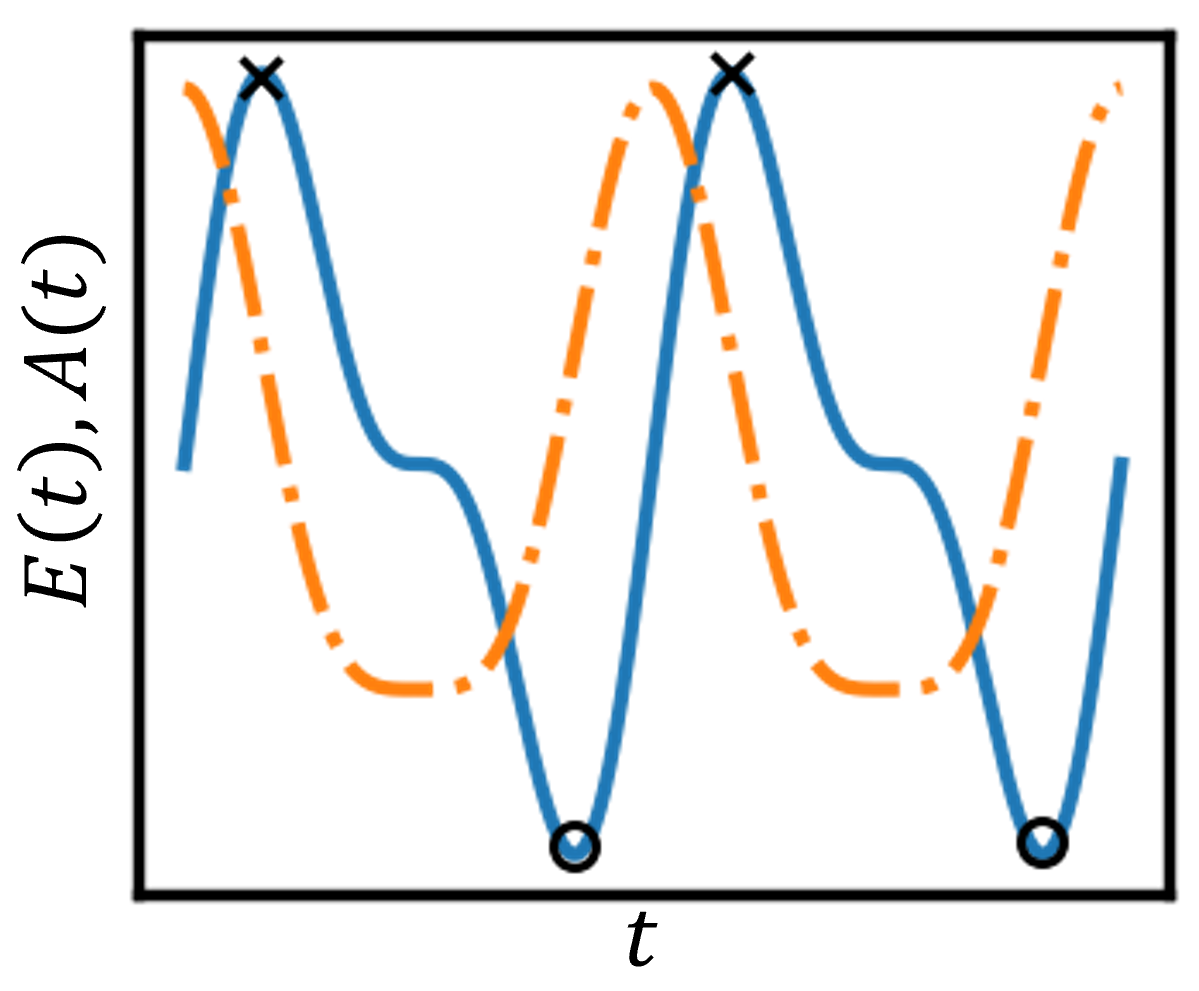}} &
\raisebox{-0.5\height}{\includegraphics[width=0.16\textwidth]{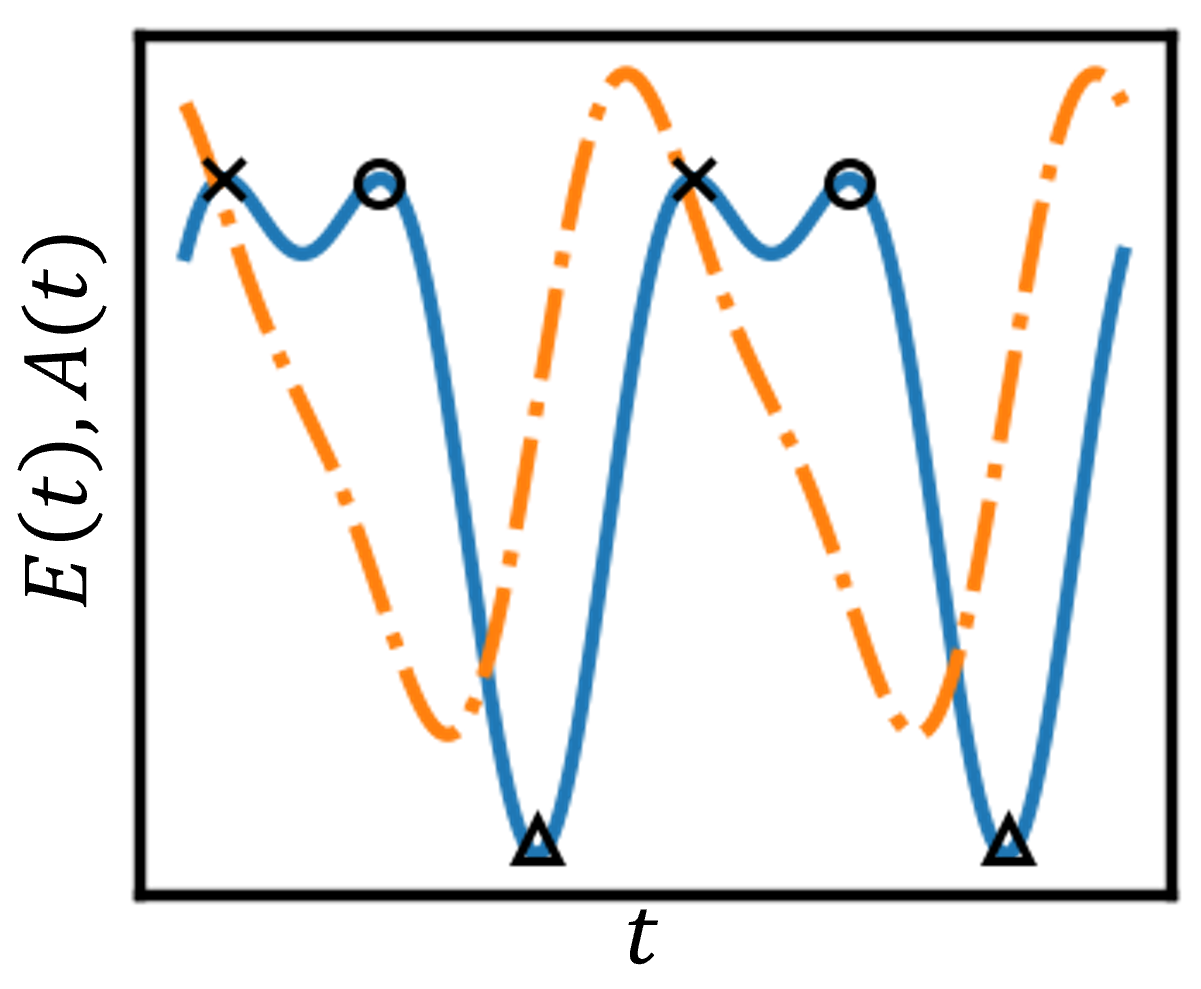}} &
\raisebox{-0.5\height}{\includegraphics[width=0.16\textwidth]{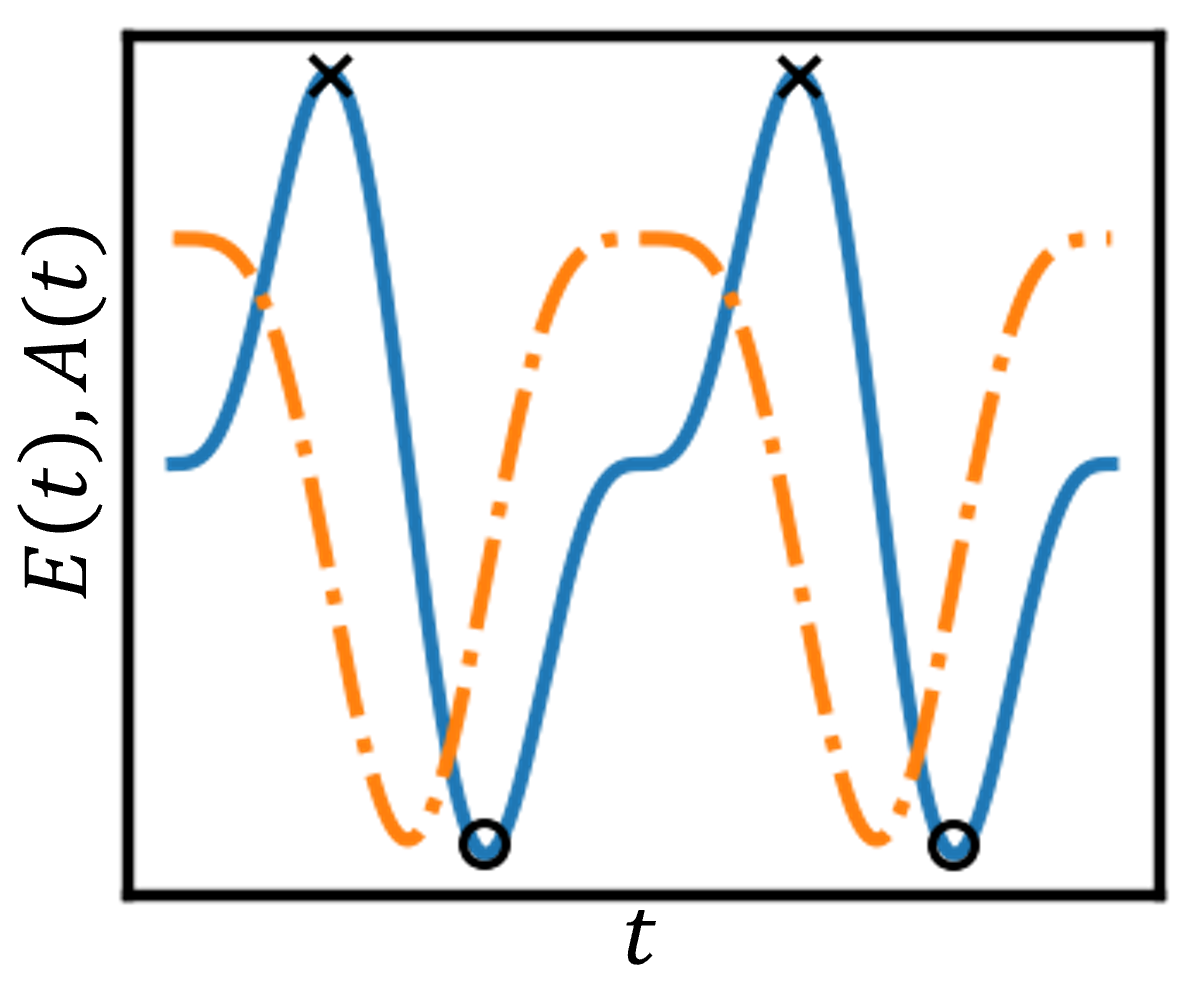}} \\
\hline
\end{tabular}
\label{tab:mapping}
\caption{Relevant ionization windows (denoted by the $\times, \bigcirc, \triangle$ and $+$ symbols)
associated with each of the events for the bichromatic driving fields displayed in
Fig.~\ref{fig:fields_overview}(c)-(c'''), from the left column to right column, respectively.
The first row gives the field parameters, the second row gives the time interval for
which these events occur, the third row states the signs of the parallel momenta
associated with each event and the final row shows a schematic of the events found
in each window.}
\end{table*}
The bichromatic field $(\omega, 3\omega, \phi=0)$ retains all symmetries of the monochromatic field as discussed in Sec.~\ref{sec:fields} and has has two ionization events per half-cycle. Hence, solutions from four `windows' are found - see corresponding schematic in the last row of Table~\ref{tab:mapping}. Because of the half-cycle symmetry, neighboring half-cycles are equivalent up to a sign change of the field [refer to corresponding second row of Table~\ref{tab:mapping}]. The times associated with the two windows in the first half-cycle are identical to those in the next half-cycle, i.e., the ionization times occur in symmetry related pairs across half-cycles, with minima offset from $p_\parallel= 0$ and increasing smoothly with the momentum. The solutions \textit{within} each half-cycle are equal to each other because the dominance of the events within each half-cycle is identical, with only the signs of the parallel momenta associated with each event differing by a sign. The PMD is symmetric around $p_{\parallel}=0$. ATI rings from intercycle interference are visible - these are now centered around the momentum values for which $\text{Im}[t]$ was minimum. The additional finer interference structure comes from interference between events in the same optical cycle. There are more such events compared to monochromatic fields, leading to more overlapping fringes.

In contrast, the $(\omega, 2\omega, \phi=0)$, $(\omega, 2\omega, \phi=\pi)$  fields [Fig.~\ref{fig:fields_overview}(c) and (c''') respectively] have one relevant event each per half-cycle, and hence two windows are utilized [see corresponding schematic in third row of Table~\ref{tab:mapping}]. The signs of the parallel momenta associated with each of these two solutions is the same as displayed in Table~\ref{tab:mapping}. Therefore the times from both windows in Fig.~\ref{fig:mlbichromatic}(b) are overlapping with each other and offset identically from $p_{\parallel}=0$. The times in Fig.~\ref{fig:mlbichromatic}(d) are identical to those in panel (b) but reflected about $p_{\parallel}=0$ since the fields are mirror images of each other too. These ideas hold for their PMDs too, which are asymmetric along the polarization axis and opposite to each other. ATI rings remain present, but are unevenly spread and sized due to the dominance of specific ionization events over others.

The $(\omega, 2\omega, \phi=\pi/2)$ field [Fig.~\ref{fig:fields_overview}(c''); fourth column of Table~\ref{tab:mapping}] has two events in one half-cycle and only one dominant event in the other half-cycle, resulting in three solutions for which three windows were utilized. The signs of the parallel momenta associated with two of the solutions are opposite, while one solution is centered around 0. Hence, the time associated with window 2 (circles) is centered at and symmetric about $p_{\parallel}=0$ whilst those associated with windows 1 and 3, are equal, opposite, and symmetric at some parallel momenta offset from zero. The PMD is also symmetric but the ATI rings are much fainter. The solutions at these peaks are associated with $\textbf{A}(t) \approx \textbf{0}$ and hence do not contribute much, leading to the PMD being smoother than that of the other fields. 
The right-most panels, showing the error demonstrates that the PINN accurately captures the evolution of the saddle times across these parameter changes, and that the PINN-guided solver converges reliably.

\subsection{Elliptically polarized and bicircular fields}\label{subsec:mlnonlinear}

\begin{figure*}
    \centering
    \includegraphics[width=\textwidth]{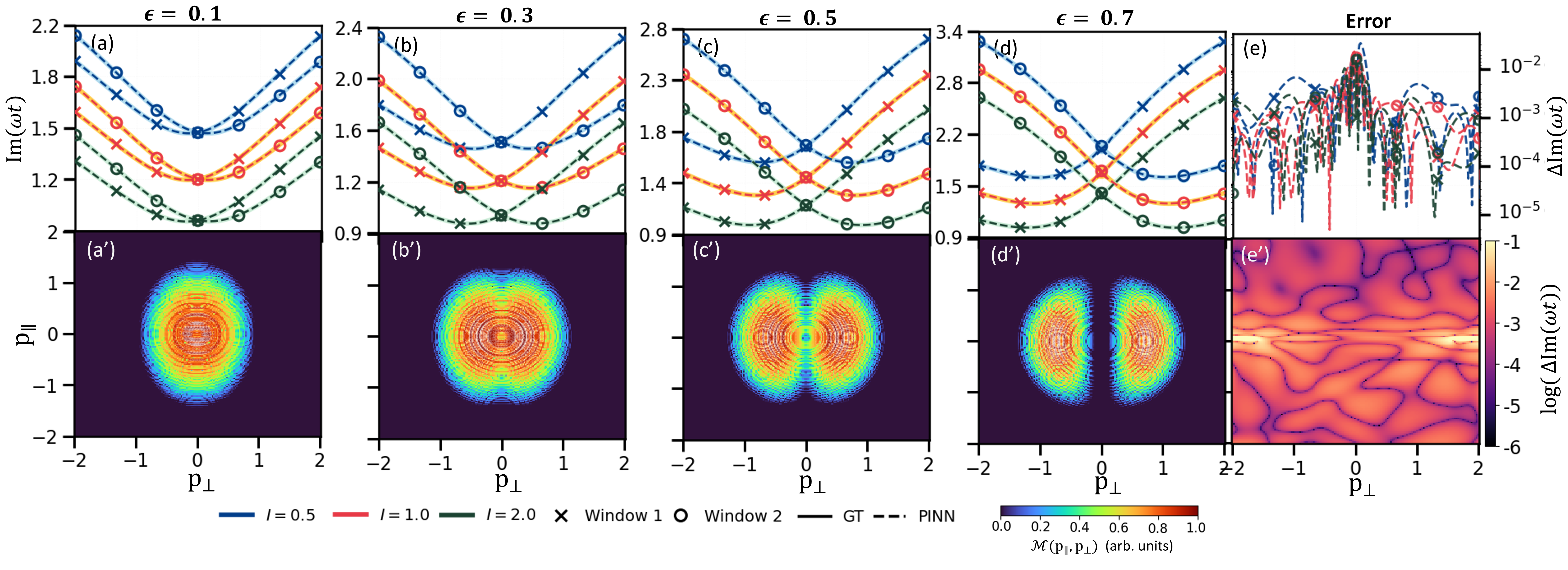}
\caption[Imaginary parts of ionization time, associated mean square error and photoelectron momentum distributions computed using the PINN solver for elliptically polarized fields with varying ellipticities]{The imaginary part of the direct ATI ionization time computed at $(p_{\parallel}, 0)$ with the PINN solver detailed in Sec.~\ref{sec:method} for the elliptically polarized fields shown in Fig.~\ref{fig:fields_overview}, [panels (a)-(d)]. For each field, multiple windows corresponding to each distinct solution are taken and denoted by crosses and circles. Three different intensities $I=0.5\times10^{14}$W/cm$^2$, $1\times10^{14}$W/cm$^2$ and $2 \times10^{14}$W/cm$^2$ have been used, denoted by blue, orange and black lines respectively. The corresponding PMDs for each field are displayed in panels (a')-(d') with $I=2\times10^{14}$W/cm$^2$ and taking all windows and multiple cycles of the fields. Panels (e) and (e') show the mean square error (MSE) in the imaginary time at $(p_{\parallel}, 0)$ and for the entire 2D momentum grid with the $\epsilon=0.3$ case, respectively. All other field parameters are the same as in the associated panels of Fig.~\ref{fig:fields_overview}.}
    \label{fig:mlelliptical}
\end{figure*}

Fig.~\ref{fig:mlelliptical} exhibits the imaginary ionization times [panels (a)-(d)] and PMDs [panels (a')-(d')] for four different ellipticities. These fields are depicted in Fig.~\ref{fig:fields_overview}(d)-(d$'''$). At small ellipticity, the field behaves approximately like its linearly polarized counterpart (the monochromatic field). There are two windows and the imaginary times associated with the each solution have minima very close to $p_\parallel =0$ and are roughly symmetric. As ellipticity increases, the minima of the solutions gets further away from $p_\parallel=0$, in opposite directions, and the solutions become less and less symmetric. The solutions are equal and occupy opposite regions of the parallel momentum space. The PMDs are reflection symmetry about $p_\parallel = - p_\parallel$ and `open up' as ellipticity increases \cite{Bashkansky1987,Kim2022}, associated with the solutions getting further away from $p_\parallel=0$. The ATI rings are visible in all distributions. The error plots in Fig.~\ref{fig:mlelliptical}(e) and (e') indicate small enough losses for the time predictions from the PINN solver to be reliable. As in the few-cycle CEP case, a single trained PINN can predict across a two-parameter space; here we predict for arbitrary pairs $(\varepsilon, I)$ without retraining, at the cost of increased training time compared to a single-parameter scan.

\begin{figure*}
    \centering
    \includegraphics[width=\textwidth]{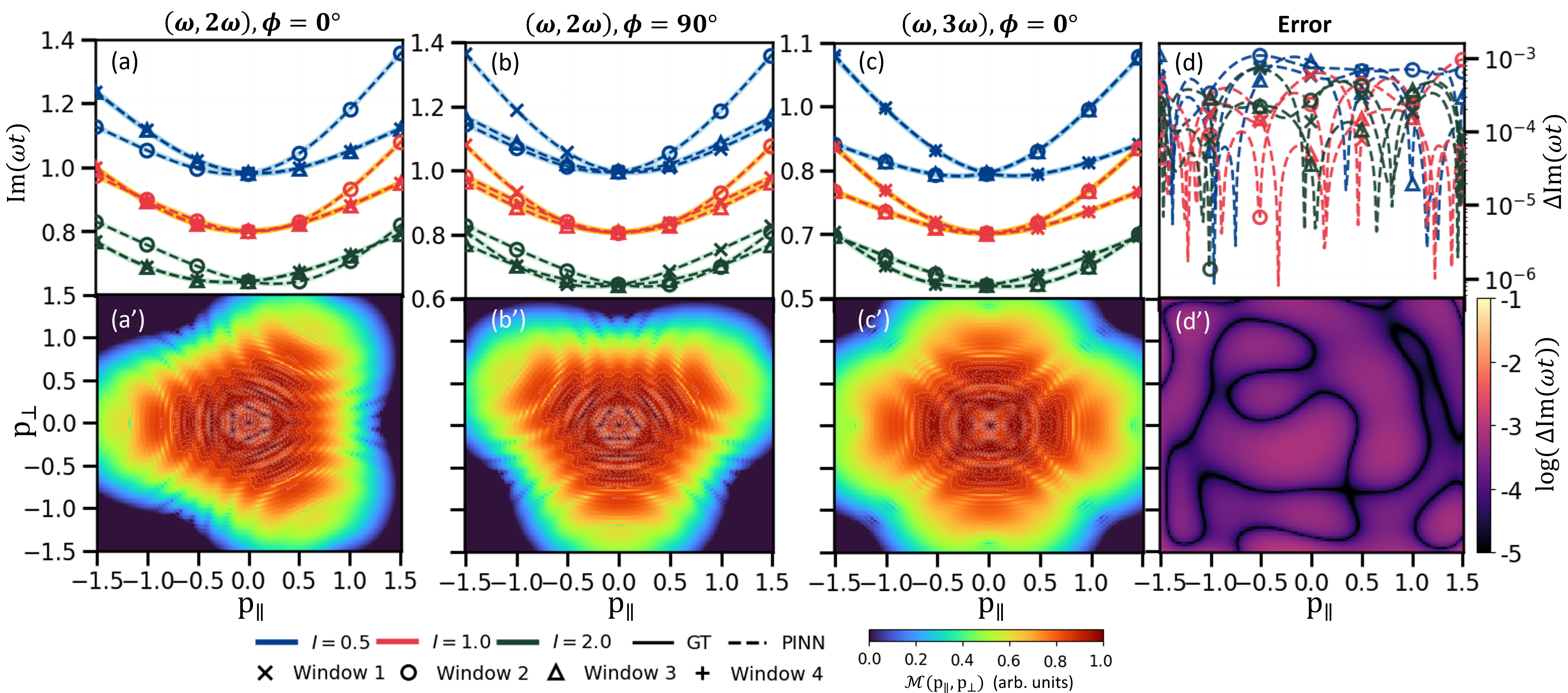}
   \caption[Imaginary parts of ionization time, associated mean square error and photoelectron momentum distributions computed using the PINN solver for bicircular fields with varying harmonic frequencies and relatives phases]{The imaginary part of the direct ATI ionization time computed at $(p_{\parallel}, 0)$ with the PINN solver detailed in Sec.~\ref{sec:method} for the bicircular fields shown in Fig.~\ref{fig:fields_overview}, [panels (a)-(c)]. For each field, multiple windows corresponding to each distinct solution are taken and denoted by crosses, circles, triangles and plus signs. Three different intensities $I=0.5\times10^{14}$W/cm$^2$, $1\times10^{14}$W/cm$^2$ and $2 \times10^{14}$W/cm$^2$ have been used, denoted by blue, orange and black lines respectively. The corresponding PMDs for each field are displayed in panels (a')-(c') with $I=2\times10^{14}$W/cm$^2$ and taking all windows and multiple cycles of the fields. Panels (d) and (d') show the mean square error (MSE) in the imaginary time at $(p_{\parallel}, 0)$ and for the entire 2D momentum grid with the $(\omega, 2\omega, \phi=2\pi)$ case, respectively. All other field parameters are the same as in the associated panels of Fig.~\ref{fig:fields_overview}}
    \label{fig:mlbicircular}
\end{figure*}

Finally, Fig.~\ref{fig:mlbicircular}(a)-(c) displays the imaginary part of the ionization time computed with the PINN solver for the bicircular fields shown in Fig.~\ref{fig:fields_overview}(e)-(e''). The solutions are computed using three (four) windows for the three-fold (four-fold) fields. 

In this case, discrete rotational symmetries of the driving field impose corresponding constraints on the ionization times and ATI distributions.
The three-fold field is invariant under rotations by $2\pi/3$ such that $E(t) \rightarrow E(t+ T/3)$. It therefore has a three-fold rotational symmetry over one optical cycle, and all three sub-cycle field extrema are equivalent. Similarly, the four-fold field is invariant under rotations by $\pi/2$ and has a four-fold rotational symmetry over an optical cycle. The PMDs [Fig.~\ref{fig:mlbicircular}(a')-(c')] inherit this symmetry directly \cite{Becker2017plane}. In each case, ATI rings from intercycle interference are visible, along with modulations from interference between events in a single cycle. When a relative phase is introduced, the periodicity of the field is preserved and hence the overall rotational symmetry in the PMD is also retained, but the distribution is rotated - compare panels (a') and (b'). The rotation of the distribution can be predicted from the imaginary times. Fig.~\ref{fig:mlbicircular}(b) associated with $(\omega, 2\omega, \phi=90^{\circ})$ shows the solutions associated with windows 1 and 2 are asymmetric about $p_\parallel=0$ and opposites of each other, with the solution associated with symmetric and having a minimum at $p_\parallel = 0$. In contrast, for $(\omega, 2\omega, \phi=0^{\circ})$ [Fig.~\ref{fig:mlbicircular}(a)], the solutions associated with windows 1 and 3 are overlapping, and are not opposite to the solution associated with window 2. The four-fold symmetry of the $(\omega, 3\omega, \phi=0^{\circ})$ case can be seen from Fig.~\ref{fig:mlbicircular}(c) too - solutions associated with windows 1 and 4 are overlapping, and equal and opposite to the solutions associated with windows 2 and 3 which are also overlapping. Whilst this appears obvious, this constitutes key evidence that the PINN solver has implicitly and reliably learnt the symmetry of the field through the window parameterization method. Once again, the error in Fig.~\ref{fig:mlbicircular}(e) and (e') is small, indicating reliable predictions from the PINN solver. 

\section{Conclusions and outlook}\label{sec:conclusions}
We have presented a physics-informed neural network framework for solving the strong-field approximation saddle-point equations relevant to direct above-threshold ionization under a broad class of tailored driving fields. Direct ATI within the SFA was chosen deliberately as a testbed: it is one of the most thoroughly understood problems in strong-field physics, with a well-characterized saddle point structure and clear known observational signatures in the PMDs. This makes it an ideal setting for a proof of concept study, allowing the performance and potential failures of the PINN solver to be assessed against a reliable reference rather than against an unresolved theoretical problem. Although PINNs have successfully been applied in other research areas, such as fluid dynamics \cite{Eivazi2022}, quantum mechanics \cite{Raissi2019,chen2025neuralpdesolversphysics}, solid mechanics \cite{Haghihat2021}, and hydrology \cite{Secci2024},
their use within strong-field physics remains limited within the context of solving SPEs. This is notable given that repeated saddle point evaluations, along with the difficulty of generalizing Newton-type root-finding algorithms to tailored fields constitute a computational bottleneck in semiclassical strong-field calculations. 
The key technical result of this work is that embedding the SPE residual directly into the training objective is sufficient to train a network that accurately recovers the complex ionization times across wide ranges of laser field parameters including intensity, carrier-envelope phase, ellipticity and polarization. Importantly, this is achieved without any labeled saddle-point solutions or prior knowledge of the dominance of relevant ionization events, requiring only the definition of the driving field and the ionization potential. These results demonstrate that unsupervised physics-informed training alone can recover the correct saddle-point structure of the problem.

Beyond the performance of the PINN solver itself, the results presented here support previous findings \cite{Habibovic2021,Neufeld2025}, which
demonstrate that direct ATI provides a transparent setting for mapping the temporal symmetries of a driving field to the structure and weighting of ionization saddle-point solutions to observable photoelectron momentum distributions.  
Our work shows that, for monochromatic fields, the presence of half-cycle symmetry and time-reflection symmetry about the extrema and zero crossings constrain the ionization times to occur in symmetric pairs, leading to symmetric PMDs elongated along the polarization axis and characterized by well-defined and centered ATI rings from inter-cycle interference. When these symmetries are selectively broken, as in bichromatic fields, the saddle-point structure changes accordingly. The number of contributing ionization events, their relative weighting, and their symmetry relationship is all modified in a predictable way, with direct consequences for the shape and symmetry of the ATI distributions. Few-cycle pulses provides a more extreme example of symmetry breaking, localizing ionization to a small number of dominant events and producing strongly CEP-dependent and asymmetric PMDs. 
For orthogonally polarized fields, increasing ellipticity causes symmetry-related saddle-point solutions to separate in the complex-time plane, progressively suppressing the interference between them and driving the ATI distributions towards ring-like structures characteristic of circular polarization. Bicircular fields introduce discrete rotational symmetries, which are faithfully imprinted onto both the saddle-point solutions and the PMDs, yielding distributions with three- and four-fold symmetry that directly reflect the symmetry of the driving field.

Moreover, these results allow us to address questions [posed in Sec.~\ref{sec:fields}] regarding the performance and robustness of the physics-informed neural network solver.
First, we consider how well the model captures the symmetries of the driving field. Across all field configurations, the PINN consistently recovers the correct number of ionization saddle points per cycle and reproduces their symmetry relationship, including equivalent solutions in half-cycles, reflection symmetry, and discrete rotational symmetry where present. This indicates that the model successfully internalizes the symmetry structure of the saddle-point equations through the physics residual alone. The solver does not introduce spurious solutions outside the physically relevant windows, nor does it miss symmetry-allowed solutions when the window size is chosen to encompass the relevant extrema, confirming that the combination of the windowing technique and unsupervised physics-based training is sufficient to recover all relevant ionization events for a range of tailored fields.

A more stringent test arises for few-cycle pulses, where the relevant question is not merely whether all possible ionization events can be found, but whether the solver can correctly identify the \textit{dominant} events as field parameters are varied. The CEP-dependent pulse results demonstrate that the PINN is indeed able to track changes in dominance between competing ionization events as the pulse shape is modified. For CEPs related by time reflection (such as CEP $65^{\circ}$ and $155^{\circ}$), the solver recovers saddle-point solutions and ATI distributions that are equal and opposite, consistent with the expected symmetry. Crucially, this behavior is obtained without labeled data, without explicit information about which events should dominate, and with only minimal adjustment of the window size. The solver effectively recognizes when the dominance of ionization events changes and responds accordingly, addressing a central limitation of conventional gradient-based root-finding methods, which often require extensive manual tuning in such scenarios.

Moreover, we have shown that the PINN accurately predicts the expected symmetry properties of the saddle-point solutions throughout multiple parameter spaces. By varying relative phases, frequency ratio, CEP, ellipticity, polarization, and intensity, we verified this holds across a broad range of parameters. The solver performs reliably in regimes where symmetries are preserved, weakly broken, or completely absent. 

Several limitations of the present framework should be noted. The architecture and the loss function were designed for direct ATI in which only the ionization SPE must be satisfied. Extensions to rescattered ATI or high-harmonic generation where an additional recollision conditions must be imposed will require modifications to the network design and a careful treatment of the additional recollision time solutions and nearly coalescent saddles. Extension to non-sequential double ionization introduces additional degrees of freedom and may require a hybrid approach in which some supervised information of physical constraints beyond the residual are incorporated into training. Similarly, moving beyond the SFA to Coulomb-distorted models such as the CQSFA introduces additional coupled equations governing the electron trajectory, the presence of additional Coulomb-distorted trajectories and the emergence of caustics and branch cuts. The appropriate PINN formulation for this model remains to be developed. We expect that each of these extensions will require non-trivial adjustments to the architecture, loss definition and potentially the training strategy. These factors represent potential directions for future work. 

Nonetheless, the results presented here establish that PINNs can serve as a practical and promising alternative to traditional root-find approaches for semi-classical strong field calculations and provide a foundation upon which more capable solvers can be built. Due to the training being unsupervised, and the ease of generalization across different tailored field families, this method (or future iterations of) could prove useful as a component within a larger optimization or solution strategies for strong-field equations, such as that informing guesstimes for Newton-type solvers. Finally, the acceleration achieved by the PINN opens the door to computationally demanding studies that require high-dimensional parameter sweeps potentially including quantum-light driving fields with tailored fields for which classical saddle-point solvers may require additional manual tuning.

\textbf{Acknowledgements: } Discussions with T. Rook, J. Wu, A. Bayat and K. Bihannic are gratefully acknowledged. This work has been funded by the UK Engineering and Physical Sciences Research Council (EPSRC) (Grants No. EP/T019530/1 - AQuADIP - and  UKRI2300 - APIQuL - Attosecond Photoelectron Imaging with Quantum Light) and by UCL. 
\raggedright
\normalem

\end{document}